\input harvmac
\font\teneufm=eufm10
\font\seveneufm=eufm7
\font\fiveeufm=eufm5
\newfam\eufmfam
\textfont\eufmfam=\teneufm
\scriptfont\eufmfam=\seveneufm
\scriptscriptfont\eufmfam=\fiveeufm

\font\teneusm=eusm10
\font\seveneusm=eusm7
\font\fiveeusm=eusm5
\newfam\eusmfam
\textfont\eusmfam=\teneusm
\scriptfont\eusmfam=\seveneusm
\scriptscriptfont\eusmfam=\fiveeusm

\font\tenmsx=msam10
\font\sevenmsx=msam7
\font\fivemsx=msam5
\font\tenmsy=msbm10
\font\sevenmsy=msbm7
\font\fivemsy=msbm5
\newfam\msafam
\newfam\msbfam
\textfont\msafam=\tenmsx  \scriptfont\msafam=\sevenmsx
  \scriptscriptfont\msafam=\fivemsx
\textfont\msbfam=\tenmsy  \scriptfont\msbfam=\sevenmsy
  \scriptscriptfont\msbfam=\fivemsy

\def\msbm#1{{\fam\msbfam\relax#1}}

\font\teneurm=eurm10
\font\seveneurm=eurm7
\font\fiveeurm=eurm5
\newfam\eurmfam
\textfont\eurmfam=\teneurm
\scriptfont\eurmfam=\seveneurm
\scriptscriptfont\eurmfam=\fiveeurm


\font\tencmmib=cmmib10  \skewchar\tencmmib='177
\font\sevencmmib=cmmib7 \skewchar\sevencmmib='177
\font\fivecmmib=cmmib5 \skewchar\fivecmmib='177
\newfam\cmmibfam
\textfont\cmmibfam=\tencmmib
\scriptfont\cmmibfam=\sevencmmib
\scriptscriptfont\cmmibfam=\fivecmmib
\def\cmmib#1{{\fam\cmmibfam\relax#1}}

\font\tenbifull=cmmib10 
\font\tenbimed=cmmib10 scaled 800
\font\tenbismall=cmmib10 scaled 666
\textfont9=\tenbifull \scriptfont9=\tenbimed
\scriptscriptfont9=\tenbismall

\def\Bz{{\fam=9{\mathchar"7124} }}

\def\a{\alpha}           
\def\c{\chi}       
\def\d{\delta}

\def\g{\gamma}          \def\k{\kappa}     
\def\l{\lambda}
\def\L{\Lambda}   \def\m{\mu}                 
\def\r{\rho}
  \def\o{\omega}      \def\O{\Omega}     
\def\p{\psi}
      \def\s{\sigma}      \def\S{\Sigma}     
\def\th{\theta}
\def\t{\tau}            

\def\z{\zeta}
\def\CA{{\cal A}}

\def\CG{{\cal G}}

\def\CK{{\cal K}}
\def\CM{{\cal M}}
\def\CN{{\cal N}}

\def\CL{{\cal L}}
\def\CD{{\cal D}}
\def\CW{{\cal W}}
 \def\CV{{\cal V}}
\def\CP{\msbm{CP}}
\def\ep{\epsilon}

\def\R{\msbm{R}}
\def\BK{\cmmib{K}}

\def\rd{\partial}

\def\darr#1{\raise1.5ex\hbox{$\leftrightarrow$}
\mkern-16.5mu #1}
\def\Ha{{1\over2}}

\def\Fr#1#2{{#1\over#2}}

\def\roughly#1{\raise.3ex\hbox{$#1$\kern-.75em
\lower1ex\hbox{$\sim$}}}

\def\opname#1{\mathop{\kern0pt{\rm #1}}\nolimits}
\def\tr{\opname{Tr}}
\def\ch{\opname{ch}}
\def\Re{\opname{Re}}
\def\Im{\opname{Im}}
\def\End{\opname{End}}
\def\pr{\prime}
\def\ppr{{\prime\prime}}
\def\bs{\cmmib{s}}
\def\bbs{\bar\cmmib{s}}
\def\Dp{\rd_{\!A}}
\def\Dpp{\bar\rd_{\!A}}
\def\mapr{\!\smash{
	    \mathop{\longrightarrow}\limits^{\bs_+}}\!}
\def\mapl{\!\smash{
	    \mathop{\longleftarrow}\limits^{\bs_-}}\!}
\def\mapbr{\!\smash{
	    \mathop{\longrightarrow}\limits^{\bbs_+}}\!}
\def\mapbl{\!\smash{
	    \mathop{\longleftarrow}\limits^{\bbs_-}}\!}
\def\mapd{\Big\downarrow
 	 \rlap{$\vcenter{\hbox{$\scriptstyle \bbs_-$}}$}}
\def\mapu{\Big\uparrow
	  \rlap{$\vcenter{\hbox{$\scriptstyle\bbs_+$}}$}}

\def\etal{et al.}
\def\git{/\kern-.25em/}
\def\cmp#1#2#3{Comm.\ Math.\ Phys.\ {{\bf #1}} {(#2)} {#3}}
\def\pl#1#2#3{Phys.\ Lett.\ {{\bf #1}} {(#2)} {#3}}
\def\np#1#2#3{Nucl.\ Phys.\ {{\bf #1}} {(#2)} {#3}}

\def\prl#1#2#3{Phys.\ Rev.\ Lett.\ {{\bf #1}} {(#2)} {#3}}

\def\jdg#1#2#3{J.\ Differ.\ Geom.\ {{\bf #1}} {(#2)} {#3}}

\def\plms#1#2#3{Proc.\ London Math.\ Soc.\ {{\bf #1}} 
{(#2)} {#3}}

\def\ihes#1#2#3{Publ.\ Math.\ I.H.E.S. \ {{\bf #1}} {(#2)} {#3}}

\def\ack{\bigbreak\bigskip\centerline{
{\bf Acknowledgements}}\nobreak}
\def\subsubsec#1{\ifnum\lastpenalty>9000\else\bigbreak\fi
  \noindent{\it #1}\par\nobreak\medskip\nobreak}

\def\lin#1{\noindent {$\underline{\hbox{\it #1}}$} }



\lref\polch{ J.~Polchinski, 
{\it Dirichlet branes and Ramond Ramond charges},  
\prl{75}{1995}{4724}, 
{\tt hep-th/9510017}\semi
J. Polchinski, S. Chaudhuri, and C. Johnson, 
{\it Notes on D-branes}, 
{\tt hep-th/9602052}\semi
J. Polchinski, {\it TASI Lectures on D-branes}, 
{\tt hep-th/9611050}.
}

\lref\bound{ E.~Witten, 
{\it Bound states of strings and $p$-branes},
Nucl. Phys.  {\bf B460} (1996) 335--350,
{\tt hep-th/9510135}.
}

\lref\SYZ{
A. Strominger, S.-T. Yau and E. Zaslow,
\np{B479}{1996}{243}.
}

\lref\Vafa{
C. Vafa, {\it Extending mirror conjecture to Calabi-Yau with bundles},
{\tt hep-th/9804131}.
}

\lref\VW{C. Vafa and E. Witten,
{\it A strong coupling test of S-Duality},
Nucl. Phys. {\bf B431} (1994) 3-77,
{\tt  hep-th/9408074}.
} 

\lref\DM{
R. Dijkgraaf and G. Moore,
{\it Balanced topological field theories},
Commun. Math. Phys. {\bf 185} (1997) 411-440,
{\tt hep-th/9608169}.
}

\lref\DPS{
R. Dijkgraaf, B.J. Schroers and J.-S. Park,
{\it N=4 supersymmetric Yang-Mills theory on a K\"{a}hler surface}, 
{\tt hep-th/9801066}.
}

\lref\Park{
J.-S.~Park,
{\it N=2 topological Yang-Mills theory
on compact K\"{a}hler surfaces},
\cmp{163}{1994}{113},
{\tt hep-th/9304060}.
}

\lref\HYM{
J.-S. Park,
{\it  Holomorphic Yang-Mills theory on compact K\"{a}hler manifolds},
Nucl. Phys. B423 (1994) 559, {\tt hep-th/9305095}
\semi
S.~Hyun and J.-S.~Park,
{\it N=2 topological Yang-Mills theories
and Donaldson's polynomials},
J. Geom. Phys. {\bf 20} (1996) 31-53, 
{\tt hep-th/9404009}.
}

\lref\BJSV{
M. Bershadsky, A. Johansen, V. Sadov and C. Vafa,
{\it Topological reduction of 4D SYM to 2D $\sigma$--models},
Nucl. Phys. {\bf B448} (1995) 166
{\tt hep-th/9501096}.
}


\lref\TFT{
E. Witten,
{\it Topological quantum field theory},
Commun. Math. Phys. {\bf 117}  (1988) 353.
}

\lref\tdYM{
E. Witten,
{\it Two dimensional gauge theories revisited},
Mod. Phys. Lett. {\bf A7} (1992) 2647,
{\tt hep-th/9204084}.
}

\lref\TSM{
E. Witten, 
{\it Topological sigma models},
Commun. Math. Phys. {\bf 118}  (1988) 411. 
}

\lref\Wittenmirror{
E. Witten, 
{\it Mirror manifolds and topological field theory},
{\tt  hep-th/9112056}.
}

\lref\Wittencss{
E. Witten,
{\it Chern-Simons gauge theory as a string theory},
{\tt hep-th/9207094}.
}

\lref\GLSM{
E. Witten
{\it Phases of $N=2$ theories in two dimensions},
Nucl. Phys. {\bf B403} (1993) 159-222,
{\tt hep-th/9301042}.
}

\lref\Wittengr{
E.~Witten, 
{\it The Verlinde algebra and the cohomology of the Grassmannian},
{\tt hep-th/9312104}.
}

\lref\Mtheory{
E. Witten, 
{\it String theory dynamics in various dimensions}, 
Nucl. Phys. {\bf B443} (1995) 85,
{\tt hep-th/9503124}.
}

\lref\chiralrings{
W. Lerche, C. Vafa and N.P. Warner,
{\it Chiral rings in N=2 superconformal theories},
Nucl. Phys. {\bf B324} (1989) 427.
}

\lref\CeVa{
S. Cecotti and C. Vafa, 
{\it Topological anti-topological fusion,} 
Nucl. Phys. {\bf B367} (1991) 359.
}

\lref\kstheory{
M. Bershadsky, S. Cecotti, H. Ooguri  and C. Vafa
{\it Kodaira-Spencer theory of gravity and exact results for 
quantum string amplitudes},
Commun. Math. Phys. {\bf165} (1994) 311-428,
{\tt hep-th/9309140}.
}

\lref\Holo{
M. Bershadsky, S. Cecotti, H. Ooguri and C. Vafa (with 
an appendix by S.Katz)
{\it Holomorphic anomalies in topological field theories},
\np{B405}{1993}{279-304}, 
{\tt  hep-th/9302103}.
}


\lref\HPA{
C.M.~Hofman and J.-S.~Park,
"Sigma Models for Quiver Variety,"
to appear.
}


\lref\Hitchin{
N.J. Hitchin,
{\it The self-duality equations on a Riemann surface},
\plms{3, 55}{1987}{59}.
}

\lref\Simpson{
C.T.~Simpson,
{\it Constructing variations of Hodge structure using Yang-Mills theory
and applications to uniformization},
J.~Amer..~Math.~Soc. {\bf 1} (1988) 867.
}

\lref\SimpsonA{
C.T.~Simpson,
{\it Higgs bundles and local systems},
\ihes{75}{1992}{5}.
}

\lref\SimpsonB{
C.T.~Simpson,
{\it Moduli of representations of the fundamental group of
a smooth projective variety}, I:\ihes{79}{1994}{47};
II:\ihes{80}{1994}{5}.
}

\lref\SimpsonC{
C.T.~Simpson,
{\it The Hodge filtration on non-Abelian cohomology},
{\tt alg-geom/9604005}.
}

\lref\SimpsonD{
C.T.~Simpson,
{\it Mixed twistor structure},
{\tt alg-geom/9705006}.
}

\lref\CVG{
N.~Chriss and V.~Ginzburg,
{\it Representation theory and complex geometry},
Birkh\"auser.
}

\lref\Kirwan{
F.~Kirwan,
{\it 
Cohomology of quotients in symplectic and algebraic geometry}, 
Math. Notes {\bf 31}. Princeton Univ. Press,
(Princeton, 1984).
}

\lref\Kobayashi{
S.~Kobayashi,
{\it 
Differential geometry of complex vector bundles},
Iwanami Shoten, Publishers and Princeton Univ. Press,
1987.
}

\lref\NJquiver{
H. Nakajima,
{\it Instantons on ALE spaces, quiver varieties and Kac-Moody
algebras},
Duke. Math. J. {\bf 76} (1994) 365.
}

\lref\DW{
R. Donagi and E. witten,
{\it Supersymmetric Yang-Mills theory and integrable systems},
{\tt hep-th/9510101}.
}

\lref\GivK{
A. Givental and B. Kim,
{\it Quantum cohomology of flag manifolds and
Toda lattices}, 
{\tt hep-th/9312096}.
}

\lref\Morrison{
D.R. Morrison,
{\it Mathematical aspects of mirror symmetry},
{\tt alg-geom/9609021}.
}

\lref\Mirror{
Essays on mirror manifolds, S.-T.~Yau ed., International Press, 
Hong Kong,1992.\semi
Mirror symmetry II, B. Greene and S.-T.~Yau eds, 
AMS/IP Studies in Adv. Math. 1., AMS, Providence, RI;
International Press, Cambridge, MA (1997).
} 

\lref\Dixon{
L.J.~Dixon,
{\it Some world-sheet results on the superpotential 
form Calabi-Yau compactifications}, 
Superstrings, Unified Theories, and Cosmology
1987, G.~Furlan et al., eds., World  Scientific, 
Singapore, 1988. pp. 67.
}

\lref\Kontsevich{
M.~Kontsevich,
{\it Homological algebra of mirror symmetry},
Proceedings of the International Congress of Mathematicians,
pp. 120, Birkh\"auser, 1995, 
{\tt alg-geom/9411018}.
}

\lref\PZ{
A.~Polishchuk and E.~Zaslow,
{\it Categorial mirror symmetry: the elliptic curve,}\
{\tt math.ag/9801119}.
}

\lref\BFSS{ T. Banks, W. Fischler, S.H. Schenker, and L. Susskind,
{\it  Theory as a matrix model: a conjecture}, 
Phys. Rev. {\bf D55} (1997) 5112-5128,
{\tt hep-th/9610043}.
}

\lref\Wati{
W.~Taylor, 
{\it D-brane field theory on compact spaces},
\pl{B394}{1997}{283},
{\tt hep-th/9611042}.
}

\lref\AGM{
P.S.~Aspinwall and B.R.~Greene, and D.R.~Morrison,
{\it Calabi-Yau moduli space, mirror manifolds and 
spacetime topology change in string theory},
\np{B416}{1994}{414}.
}

\lref\Vafac{
C.~Vafa,
{\it c-Theorem and the topology of 2-d QFTs},
\pl{B212}{1988}{28}.
}

\lref\Atiyah{
M. Atiyah,
{\it Convexity and commuting Hamiltonians},
Bull.~London Math.~Soc. {\bf 14} (1982) 1.
}

\lref\AB{
M.~Atiyah and R.~Bott,
{\it The momentum map and equivariant cohomology},
Topology {\bf 23} (1984) 1.
}

\lref\WittenMorse{
E.~Witten,
{\it Supersymmetry and Morse theory},
\jdg{17}{1982}{692}.
}

\lref\NJhs{
H.~Nakajima,
{\it Lectures on Hilbert schemes of points on surface},
Preprint.
}

\lref\Kaledin{
D.~Kaledin,
{\it Hyperk\"ahler structures on total spaces of
holomorphic cotangent bundles},
{\tt alg-geom/9710026}.
}

\lref\Hausel{
T.~Hausel,
{\it Compactification of moduli of Higgs bundles},
{\tt math.AG/9804083}.
}

\lref\Donaldson{
S.K.~Donaldson,
{\it Infinite determinants, stable bundles and curvature},
Duke Math.~{\bf 54} (1987) 231.
}

\lref\UY{
K.K.~Uhlenbeck and S.T.~Yau,
{\it The existence of Hermitian Yang-Mills connections on stable
bundles over K\"{a}hler manifolds}.
}

\lref\SV{
A. Strominger and C. Vafa,
{\it Macroscopic origin of the Bekenstein-Hawking entropy},
\pl{B379}{1996}{99}, {\tt hep-th/9601029}.
}

\lref\fivebrane{ 
R.~Dijkgraaf, E.~Verlinde and H.~Verlinde, 
{\it BPS spectrum of the five-brane and black hole entropy}, 
Nucl. Phys.  {\bf B486} (1997) 77, 
{\tt hep-th/9603126}; 
{\it BPS quantization of the five-brane},
Nucl. Phys. {\bf B486} (1997) 89, 
{\tt hep-th/9604055}.
}

\lref\mbh{
R. Dijkgraaf, E. Verlinde, and H. Verlinde, 
{\it 5D black holes and matrix strings},
\np{B506}{1997}{121},
{\tt hep-th/9704018}.
}

\lref\DVVrev{
R. Dijkgraaf, E. Verlinde, H. Verlinde,
{\it Notes on matrix and micro strings},
{\tt hep-th/9709107}.
}

\lref\JuanLN{
J.M. Maldacena
{\it The large N limit of superconformal field theories and 
supergravity},
Adv. Theor. Math. Phys {\bf 2} (1998) 231,
{\tt hep-th/9711200}.
}

\lref\HPC{
C.~Hofman and J.-S.~Park,
{\it Some comments on Holomorphic Yang-Mills theory},
in preparation.
}

\lref\HP{
C.~Hofman and J.-S.~Park,
{\it Sigma models for quiver variety},
to appear.
}

\lref\Robbert{
R. Dijkgraaf,
{\it Instanton strings and HyperK\"{a}hler geometry},
\np{BN543}{1999}{544},
{\tt hep-th/9810210}.
}

\lref\NMVW{
J.A.~Minahan, D, Nemeschansky, C.~Vafa and N.P.~Warner,
{\it E-strings and $N=4$ topological Yang-Mills theory}, 
{\tt hep-th/9802168}.
}

\lref\ABKSS{
O. Aharony, M. Berkooz, S. Kachru, N. Seiberg and E. Silverstein,
{\it Matrix description of interacting theories in six dimensions},
Adv. Theor. Math. Phys.{\bf 1} (1998) 148-157,
{\tt hep-th/9707079}.
}

\lref\ABS{
O. Aharony, M. Berkooz and N. Seiberg, 
{\it Light-cone description of (2,0) superconformal theories 
in six dimensions}, 
Adv. Theor. Math. Phys. {\bf 2} (1998) 119-153,
{\tt hep-th/9707079}.
}

\lref\GMP{
B. Greene, D. Morrison, M. Plesser,
{\it Mirror manifolds in higher dimension},
\cmp{173}{1995}{559},
{\tt  hep-th/9402119}.
}

\lref\orbifolds{
 R. Dijkgraaf, G, Moore, E. Verlinde, and H. Verlinde,
{\it Elliptic genera of symmetric products and second quantized strings},
Commun. Math. Phys. {\bf 185} (1997) 197,
{\tt hep-th/9608096}.
}

\lref\motl{
L.~Motl, 
{\it Proposals on non-perturbative superstring interactions},
{\tt hep-th/9701025}.
}

\lref\tomnati{
T.~Banks and N.~Seiberg, 
{\it Strings from matrices},
{\tt hep-th/9702187}.
}

\lref\matrixstring{ 
R. Dijkgraaf, E. Verlinde, and H. Verlinde, 
{\it Matrix string theory}, 
Nucl. Phys. {\bf B500} (1997) 43, 
{\tt hep-th/9703030}.
}

\lref\deBor{
J. de Boer,
{\it Six-dimensional supergravity on $S^3 X AdS_3$ and 
2d conformal field theory},
\np{B548}{1999}{139},
{\tt hep-th/9806104}\semi
{\it Large N elliptic genus and AdS/CFT correspondence},
JHEP {\bf 9905} (1999) 017,
{\tt hep-th/9812240}.
} 

\lref\HarveyMoore{
J. Harvey and G. Moore,
{\it One the algebras of BPS states},
Commun. Math. Phys. {\bf 197} (1998) 489
{\tt hep-th/9609017}.
}

\lref\Inflow{
M. Green, J. harvey and G. Moore,
{\it I-brane inflow and anomalous couplings on D-branes}, 
Class. Quant. Grav. {\bf 14} (1997) 47,
{\tt hep-th/9605033}.
}

\lref\Thomas{
R.P. Thomas,
{\it A holomorphic Casson invariant for Calabi-Yau $3$-folds, and
bundles on $K3$ fibrations},
{\tt math.AG/9806111}.
}

\lref\ThomasTh{
R.P. Thomas,
{\it Gauge Theory on Calabi-Yau manifolds},
PhD Thesis.
}

\lref\MorP{
D.R. Morrison, M.R. Plesser
{\it Summing the instantons: Quantum cohomology and 
mirror symmetry in toric varieties},
Nucl. Phys. {\bf B440} (1995) 279-354,
{\tt hep-th/9412236}.
}

\lref\Why{
N. Seiberg,
{\it Why is the matrix model correct?},
Phys. Rev. Lett. {\bf 79} (1997) 3577-3580,
{\tt hep-th/9710009}.
}

\lref\Sen{
A. Sen,
{\it D0 branes on $T^n$ and matrix theory}, 
Adv. Theor. Math. Phys. {\bf  2} (1998) 51,
{\tt hep-th/9709220}.
}


\lref\Douglas{
M.R. Douglas,
{\it D-branes in curved space},
{\tt hep-th/9703056}\semi
{\it D-branes and matrix theory in curved space},
lecture given at  Strings '97,
Nucl. Phys. Proc. Suppl. {\bf 68} (1998) 381,
{\tt hep-th/9707228}.
}

\lref\DKO{
M.R. Douglas, A. Kato and H. Ooguri,
{\it D-brane actions on K\"{a}hler manifolds},
{\tt hep-th/9708012}.
}

\lref\Wynter{
Ph. Brax, T. Wynter,
{\it Limits of matrix theory in curved space},
\np{B546}{1999}{182},
{\tt hep-th/9806176}.
}

\lref\Tyurin{
A. Tyurin, 
{\it Geometric quantization and mirror symmetry},
{\tt math/9902027}.
}


\lref\Nekras{N. Nekrasov, PhD. Thesis, Princeton 1996}

\lref\Baulieu{L. Baulieu, A. Losev and N. Nekrasov, 
{\it Chern-Simons and Twisted Supersymmetry in Various Dimensions},
\np{B522}{1998}{82},
{\tt hep-th/9707174}.
}


\font\Titlerm=cmr10 scaled\magstep3

\pageno=0
\nopagenumbers
\rightline{
SPIN-99/8, ITFA-99-4, hep-th/9904150
}

\vskip .5in

\centerline{\fam0\Titlerm 
Sigma Models for Bundles on Calabi-Yau:}
\vskip .2in
\centerline{\fam0\Titlerm 
A Proposal for Matrix String Compactifications}

\tenpoint\vskip .4in
\centerline{
Christiaan Hofman$^{1}$
and
Jae-Suk Park$^{1,2}$ 
}
\vskip .2in

\centerline{\it ${}^1$Spinoza Institute}
\centerline{\it University of Utrecht}
\centerline{\it Leuvenlaan 4, 3508 TA Utrecht}

\centerline{and}

\centerline{\it ${}^2$Institute for Theoretical Physics}
\centerline{\it University of Amsterdam}
\centerline{\it Valckenierstraat 65, 1018 XE Amsterdam}
\centerline{\it The Netherlands}
\vskip .3in

\abstractfont

\noindent
We describe a class of supersymmetric gauged linear sigma-model, 
whose target space is the infinite dimensional 
space of bundles on a Calabi-Yau $3$- or $2$-fold. 
This target space can be considered the configuration space of 
D-branes wrapped around the Calabi-Yau. We propose 
that this model can be used to define matrix string theory 
compactifications. In the infrared limit the model flows to 
a superconformal non-linear sigma-model whose target space is 
the moduli space of BPS configurations of branes on the compact 
space, containing the moduli space of semi-stable bundles.
We argue that the bulk degrees of freedom decouple in the infrared 
limit if semi-stability implies stability. We study topological 
versions of the model on Calabi-Yau $3$-folds. The resulting 
$B$-model is argued to be equivalent to the holomorphic Chern-Simons 
theory proposed by Witten. The $A$-model and half-twisted model 
define the quantum cohomology ring and the elliptic genus, 
respectively, of the moduli space of stable bundles on a 
Calabi-Yau $3$-fold.

\vskip 0.3in
\Date{April, 1999}

\tenpoint
\baselineskip=16pt plus 2pt minus 1pt

\newsec{Introduction}

It is by now well established that non-perturbative string theory 
fits into a greater scheme, involving also 11 dimensional supergravity, 
which we call M-theory \Mtheory. A full 
microscopic foundation for this theory is however still lacking.  
In the matrix theory proposal of \BFSS, the full dynamics of 
uncompactified M-theory was proposed to be captured by 
a certain large $N$ limit of supersymmetric 
matrix quantum mechanics. This matrix quantum mechanics 
arises as the quantum theory of many partons, 
which are the only degrees of freedom left in the 
infinite momentum frame in the uncompactified situation. 
The partons can be identified with the D-particles in 
the corresponding type IIA string theory. When the 
theory is compactified on a circle, this leads 
to matrix string theory \matrixstring, which is described 
by the maximal $\CN_{ws}=(8,8)$ supersymmetric $U(N)$ Yang-Mills 
theory in 1+1 dimensions. In the infrared, this theory describes 
the ordinary string moving on a symmetric 
product of the transverse target space $\R^8$. 

A full description of non-perturbative 
string theory crucially involves D-branes \polch.  
D-branes are effectively described by (supersymmetric) 
gauge theories, living on the worldvolume. 
Even general configurations of D-branes are described by 
gauge theories involving non-trivial gauge configurations; 
the different branes are described by the fluxes in the 
gauge theory \bound. Configurations of D-branes may therefore 
be viewed as a stringy description of vector bundles 
(or more generally sheaves) \HarveyMoore.
Matrix (string) theory compactified on a 
non-trivial manifold should involve also the 
degrees of freedom for branes wrapped around cycles 
in the compactification manifold. Indeed, it was already 
shown in \BFSS\ that membranes could be described 
in the original M(atrix) theory. The simplest 
compactification manifolds are tori. The compactified
M(atrix) theory is described by a supersymmetric gauge 
theory living on the dual torus \BFSS\Wati. For the 
circle, this leads to the matrix string theory. 
For compactifications on a higher dimensional torus 
$S^1\times T^n$, we may view the full gauge theory 
on the dual torus $\widehat S^1\times\widehat T^n$
from a matrix string perspective as a gauged 
linear sigma model whose target space is the 
infinite dimensional space provided by the 
gauge theory on the torus $\widehat T^n$. 
The covariant derivatives on the dual torus are 
identified with (some of) the adjoint scalars 
that live on the world-sheet, and which are now 
infinite dimensional matrices. In this sense, 
it is a sector of the large $N$ matrix string theory. 
These gauge theories automatically describe the 
dynamics of the wrapped D-branes on the torus, 
which are represented by the fluxes in this gauge theory. 
We may even view the gauge theory on the dual torus 
as the  configuration space of the wrapped D-branes.

In this paper we propose to describe 
compactifications of the matrix string 
related to more general Calabi-Yau manifolds  
by gauged linear sigma models whose target space 
is the infinite dimensional space of gauge bundles. 
As for the torus, we may view this linear target space as 
the fibre of an infinite rank gauge bundle on the worldsheet. 
Also the gauge group is infinite dimensional, and is 
formed by the gauge transformations in the bundle on 
the Calabi-Yau manifold. 
In general, not all the supersymmetry will 
be preserved. For example for compactifications 
on $K3$ and $CY_3$, we should have 
$\CN_{ws}=(4,4)$ and $\CN_{ws}=(2,2)$ supersymmetry 
respectively. Note that in general the 
target space not only is described by the pure gauge bundles, 
but also include certain (adjoint) scalars, which describe 
the movement of branes in the bulk. We should remark that 
such a model is not directly related to matrix string 
theory compactified on the Calabi-Yau manifold under 
consideration. Indeed, for the torus we know that the 
gauge theory lives on the dual torus. Therefore, 
this model should more appropriately be considered 
as the matrix string compactified on some dual 
manifold. This dual manifold should be a certain 
moduli space of bundles. For example, the dual torus 
can be considered the moduli space flat bundles on 
the torus, while also for the compactification 
on $K3$ surfaces the compactification manifold of the 
matrix string is a dual $K3$, which is identified as a 
certain moduli space of bundles on the $K3$ space where 
the gauge bundles live. 

Formally the infrared limit of the matrix string 
corresponds to the limit where the bulk string coupling
constant becomes zero. The theory then flows to a 
superconformal non-linear sigma model, whose target space is 
the locus of vanishing potential; this is the moduli space of 
the gauge bundles describing the linear sigma model. 
This target space 
should then be identified with the space on which the 
fundamental string lives -- or rather a symmetric product 
of it, as the matrix string described second 
quantized string theory. Indeed, this was found for 
the uncompactified matrix string \matrixstring. 
Also for compactifications on four dimensional 
Calabi-Yau manifolds $K3$ and $T^4$, it is known that 
the appropriate moduli space of bundles is related to 
the symmetric product of the manifold. This symmetric 
product is in general smoothed out, as the resolution 
of the quotient singularity is a marginal deformation 
of the matrix string theory. For higher dimensional 
compactifications, the interaction of the strings 
corresponds to an irrelevant operator. Therefore, 
we do not expect the infrared target space to be 
given by a symmetric product in any limit in 
parameter space. For Calabi-Yau 3-folds, 
we do not know of any relation in general between 
the infrared target space of our proposed model 
and a symmetric product (it may however be 
that asymptotically for well separated strings 
this space looks like a symmetric product). 
For certain special cases however where the Calabi-Yau 
manifold is a $K3$-fibration and the dimension 
of the infrared target space is exactly 6, it 
is known that this space is a Calabi-Yau manifold 
\Thomas. But certainly, we should at least find a finite 
dimensional infrared target space, if we want to make 
sense out of this theory. This already puts very strong 
conditions on the model, and seems to imply that we can 
not define our model beyond the Calabi-Yau case. 
Even if the infrared target space is not a symmetric 
product, we may still identify our model as a 
compactification of matrix string theory, but in a more 
generalized sense. The only thing we can not do is 
the identification of the usual string theory in a 
geometrical way. 

Apart from a proposal for matrix string compactification on 
Calabi-Yau manifolds, the model we describe in this paper 
can also be considered as a natural scheme to study 
topological properties of stable holomorphic bundles 
on  Calabi-Yau manifolds. Because of the relation between 
these bundles and BPS configurations of branes, we could also 
physically view this as a model studying these BPS 
states. Natural elements to study are counting formulae, 
which have natural physical interpretations as black hole 
entropies, and (quantum) intersection rings of these 
configuration spaces. These calculations will unfortunately 
be outside the scope of this paper, although we make a start 
by studying certain properties of the topologically 
twisted models. We hope to come back to these interesting 
properties in the future.

The paper is organized as follows. In the section 2, we argue 
why gauged linear sigma models are the natural setting to study 
matrix string theory compactifications. We then give an overview 
of the general gauged linear sigma approach that we will be using. 
We will use for this the language of equivariant cohomology, rather 
than the more standard superspace approach, so this part can also 
be considered as an introduction of our notations. we also comment on 
possible relations with the non-linear sigma model approach 
proposed by Douglas \etal\ \Douglas\DKO. 

In section 3, we introduce the actual model 
describing bundles on a Calabi-Yau $3$-fold. 
we start with describing some properties of the space of 
bundles which we need for the formulation of the model. After 
that, the construction of the model will be straightforward. 
We then study the infrared limit of the theory, described by a 
non-linear sigma model. Then we study the case where the Calabi-Yau 
$3$-fold is of the form $K3\times T^2$. This relates the 
model to the matrix string description of the five-brane \DVVrev. 
We conclude this section by studying the decoupling from bulk 
degrees of freedom. 

In section 4, we consider topologically twisted versions 
of the model, along the lines of \TSM\Wittenmirror. 
The localization and observables in the $A$- and $B$-model are studied.

\newsec{Preliminaries}

In this section we review some salient features of 
$\CN_{ws}=(2,2)$ gauged linear sigma models (GLSM) \GLSM\Wittengr, 
in a language suitable for our purpose. This also involves  an 
infinite dimensional generalization of the usual GLSM, 
which can be described in the language of equivariant cohomology. 
We also briefly compare our proposal the one of Douglas \etal, 
in terms of non linear sigma models.

\subsec{$\CN_{ws}=(2,2)$ Gauged Linear Sigma Model 
and Equivariant Dolbeault Cohomology
}

We shall now describe the gauged linear sigma models (GLSM) in some more 
detail, but still in a quite general sense. The GLSM's we describe 
are slightly generalized, as we allow for an equivariant extension of 
the supersymmetry; that is we allow the supersymmetry algebra to be closed 
up to certain gauge transformations. Also, we want to generalize to 
allow for infinite dimensional target spaces. We will concentrate on 
theories which have $\CN_{ws}=(2,2)$ worldsheet supersymmetry, as this 
is the amount of supersymmetry expected for Calabi-Yau 3-fold 
compactifications. This amount of supersymmetry generally requires a 
target space which allows for a K\"ahler structure. Furthermore, 
as we want to get a linear sigma model, the target space will generally 
be flat. To have an anomaly free theory, we will also require that the 
first Chern class of the K\"ahler manifold vanishes. 
So we consider a flat K\"{a}hler manifold, which we denote 
$\CA$. The path integral of the sigma model on $\S$
with target space $\CA$ involves the space of all maps
$A: \S\rightarrow \CA$. Because of the K\"ahler structure, we can split 
up these coordinates into complex coordinates $A^i$ and their complex 
conjugates $A^{\bar\imath}$. 
The left and right super-partners $\p^i_\pm$ of the two dimensional 
scalars $A^i$ are spinors on the worldsheet
and  holomorphic tangent vectors in the target space. The $\pm$ indices 
will denote worldsheet spinor indices. When we restrict to a point in $\S$, 
we see from this and the supersymmetry commutation relations (which are 
trivial on a point) that the left and right supercharges act as 
two copies $d_\pm$ of the exterior derivative $d$ on $\CA$. 
In terms of field theory, we may also state this as the relation that the 
supersymmetry restricted to a point on $\S$ reduces to a BRST 
symmetry on the target space $\CA$, as $d^2=0$. 
The K\"ahler structure on the target space $\CA$ implies that we have 
$\CN_{ws}=(2,2)$ supersymmetry, due to the decomposition $d=\rd +\bar\rd$ 
of the exterior derivative. 

Now we consider the case that a group $\CG$ acts on $\CA$
preserving the complex and K\"{a}hler structures.
One may attempt to define a sigma model for the quotient space 
$\CA/\CG$ by gauging the symmetry $\CG$. The problems one may  
encounter is that we rarely have a good quotient and the 
K\"{a}hler structure may not descend to it.
The $\CN_{ws}=(2,2)$ gauged sigma model however resolves these 
problems in whole sale! To describe it, we use the relation between 
the supersymmetry on the worldsheet and the BRST symmetry 
(exterior derivative) in the target space noted above. 
The BRST cohomology is naturally generalized in the gauged 
situation to so called equivariant cohomology. 
The hart of the construction is an {\it automatic}
equivariant extension of the space $\CA$ to $\CA_\CG 
=E\CG\times_\CG \CA$ where $\CG$ acts {\it freely}. Here 
$E\CG$ denotes the universal $\CG$-bundle.\foot{
For details on equivariant cohomology the reader is referred 
to the papers \AB\tdYM. We will always use the Cartan model.}
The left and right supercharges now act 
on $\CA$ as two copies of $\CG$-equivariant exterior derivatives
(the exterior derivatives in $\CA_\CG$), which satisfy the modified 
commutation relations 
\eqn\paa{
d_\CG=d -i \phi^a i_a, \qquad  d_\CG^2 = -i\phi^a\CL_a
}
where $\phi^a$ denotes the generator of the $\CG$-action, 
$i_a$ denotes the contraction with the vector field $V^a$ associated
with the $\CG$-action and $\CL_a$ is the Lie-derivative with respect 
to this vector field. The $\CG$-equivariant cohomology of $\CA$
is the ordinary cohomology of the extended space $\CA_\CG$.
The super-partners of $A$
become equivariant differential one forms on $\CA$.
As for the ordinary derivative $d$ we have a decomposition 
of the equivariant derivative $d_\CG$ as 
$d_\CG =\rd_\CG +\bar\rd_\CG$, such that
\eqn\pab{
\rd_\CG^2=0,\qquad \{\rd_\CG,\bar\rd_\CG\}=-i\phi^a \CL_a,
\qquad \bar\rd_\CG^2=0,
}
which defines equivariant Dolbeault cohomology \Park. 

Such a decomposition again implies an extension to extended 
$\CN_{ws}=(2,2)$ worldsheet supersymmetry, using the 
equivariant derivatives above to construct the worldsheet 
supersymmetry generators. We denote the $\CN_{ws}=(1,1)$ 
supercharges by $Q_\pm=\bs_\pm +\bbs_\pm$, 
where $\pm$ denotes the left and right spinor indices
on the worldsheet $\S$. If we reduce $\S$ to a point then
$Q_\pm$ are two copies of the equivariant exterior derivative 
$d_\CG$ and $\pm$ denote the charges under an internal symmetry of
a graded equivariant cohomology. Such 
graded equivariant differentials first appeared as twisted 
supercharges of four-dimensional $\CN=4$ SYM \VW, 
and in general are called  balanced equivariant 
differentials \DM. 
The further decomposition $Q_\pm\to\bs_\pm\oplus\bbs_\pm$
in the K\"{a}hler case gives rise to differentials of a balanced 
equivariant Dolbeault cohomology \DPS\ on the target space. 
The supercharges satisfy the following commutation 
relations following \pab
\eqn\aaa{
\eqalign{
\{\bs_+,\bbs_+\} &= i\nabla_{\!++}  ,\cr
\{\bs_-,\bbs_-\} &= i\nabla_{\!--},\cr
}\qquad
\eqalign{
\{\bs_+,\bbs_-\}&=-ig_s^{-1}\s^a\CL_a,\cr
\{\bbs_+,\bs_-\}&=-ig_s^{-1}\bar\s^a\CL_a,
}
\qquad
\eqalign{
\{\bs_+,\bs_-\} &= 0,\cr
\{\bbs_+,\bbs_-\} &= 0,
}\qquad
\eqalign{
\bs_\pm^2=0,\cr
\bbs_\pm^2=0,
}
}
where $\nabla_{\!\pm\pm}=\rd_{\pm\pm}-v_{\pm\pm}^a\CL_a$ are the covariant 
derivatives on the worldsheet $\S$
and $g_s$ is the string coupling constant, which has scaling dimension 
one on the worldsheet $\S$.
Here we have introduced gauge fields $v_{\pm\pm}$ on $\S$ and 
the group $\CG$ is extended to a group of local gauge transformations 
on $\S$. Note that $\s$ is an adjoint scalar for this gauge group.\foot{Note 
that the gauge field $v_{\pm\pm}$ is anti-hermitian in this convention, 
while also $\bar\s=-\s^\dagger$.} We see that in effect the 
equivariant extension leads on the worldsheet to a 
gauging of the symmetry by $\CG$. We should note that the above 
supersymmetry algebra is the dimensional reduction of the $\CN=1$ 
supersymmetry in 4 dimensions, where the fields $\s$ and $\bar\s$ 
are the reduced components of the gauge field \GLSM\Wittengr.
We may interpret these supercharges as differentials of a balanced
$\CG\times P_\S$-equivariant cohomology, where $P_\S$
denotes the group of translations along $\S$.
The internal consistency of the commutation relations \aaa\ 
determines a $\CN_{ws}=(2,2)$ vector multiplet, transforming according 
to the diagram 
\eqn\pad{

\matrix{
\bar\s     & \mapr & \eta_+ & \mapl & v_{++}     \cr
\mapd      &       & \mapd  &       & \mapd      \cr
\bar\eta_- & \mapr & D      & \mapl & \bar\eta_+ \cr
\mapu      &       & \mapu  &       & \mapu      \cr
v_{--}     & \mapr & \eta_- & \mapl & \s         \cr
}
}
The supersymmetry above and this vector multiplet can be found 
also from dimensional reduction of the four dimensional $\CN=1$ 
supersymmetry and vector multiplet. The complex scalar $(\s,\bar\s)$ 
then corresponds to the components along the compactified directions. 
Physically, the extension of $\CA$ to $\CA_\CG$ is just 
gauging of the global symmetry $\CG$ of the target space $\CA$.
The supermultiplet associated with the bosonic field $A^I(x)$
is completely determined by the complex structure on $\CA$.
Decomposing $A^I = A^i + A^{\bar\imath}$ as earlier
the $A^i$ should extend to a chiral (holomorphic) multiplet, 
i.e. $\bbs_\pm A^i=0$
\eqn\pae{

\matrix{
\p^i_- & \mapl & A^i & \mapr & \p^i_+ \cr
  & \rlap{\lower.3ex\hbox{$\scriptstyle \bs_{\!+}$}}\searrow
 && \swarrow\!\!\!\rlap{\lower.3ex\hbox{$\scriptstyle \bs_{\!-}$}} & \cr
&& H^i &&
}.
}

An important property of the $\CN_{ws}=(2,2)$ supersymmetric model 
is the $U(1)$ $\CR$-symmetry.
The left and right $\CR$-charges $(J_L,J_R)$ of the 
supercharges are set to the following values 
\eqn\rcharge{\eqalign{
&\bs_+ : (+1,0),\qquad \bbs_+ : (-1,0),\cr
&\bs_- : (0,+1),\qquad \bbs_- : (0,-1).
}
}
The $\CR$-charges of the fields in the vector multiplet are then 
determined by the assignment of zero charges to the gauge field; 
this gives 

\lin{Table 1}
$$

\matrix{
    & v_{++} & v_{--} & \s & \bar\s & D & \eta_+ & \eta_- & \bar\eta_+ & \bar\eta_-
\cr
J_L & 0      & 0      & +1 & -1     & 0 & 0      & +1     & 0          & -1
\cr
J_R & 0      & 0      & -1 & +1     & 0 & +1     & 0      & -1         & 0   
}
$$
Together with the obvious left and
right spin charges they determine the graded form degrees of
balanced equivariant Dolbeault cohomology.

The action functional of the theory is defined by
\eqn\paf{
\eqalign{
S(r,\bar r)= 
&\bs_+\bs_-\bbs_+\bbs_-\!\int_\S \!d^2\!x\,\Bigl(
-\tr(\s\bar\s) + \CK\bigl(A^i, A^{\bar\imath}\bigr)  
\Bigr)
\cr 
&
+ \Fr{r}{g_s} \bbs_+\bs_-\!\int \!d^2\!x\, \tr\s
+\Fr{\bar r}{g_s} \bs_+\bbs_-\!\int \!d^2\!x\, \tr\bar\s
\cr
&
+\Fr{1}{g_s}\bs_+\bs_- \!\int \!d^2\!x\, \CW\bigl(A^i\bigr)
+\Fr{1}{g_s}\bbs_+\bbs_-\!\int \!d^2\!x\,\overline\CW\bigl(A^{\bar\imath}\bigr),
}
}
where $\CK(A^i,A^{\bar\imath})$ denotes the K\"{a}hler potential
of the flat space $\CA$ and $r=i\zeta +\th/2\pi$ belongs to the center
of $Lie(\CG)\approx Lie(\CG)^*$. The trace is some suitable trace 
in a representation of the gauge group $\CG$. 
The action functional $S$ is obviously invariant 
under the $\CN_{ws}=(2,2)$ the supersymmetry, since
$(\bbs_+\oplus\bbs_-)A^i=(\bs_+\oplus\bbs_-)\s=0$.

{}From \pad\ we see that the generators of balanced equivariant Dolbeault 
cohomology consist of the bosonic fields in a $\CN_{ws}=(2,2)$ vector multiplet.
The remaining  bosonic auxiliary fields $D$ and $H_i$ form a crucial 
ingredient of the theory. Imposing the {\it algebraic}  equation of motions 
for these fields one always has 
\eqn\pag{
\eqalign{
D   &= \Fr{1}{g_s^2}\left(\m - \zeta\right),\cr
H_i &= \Fr{1}{g_s}\left(\Fr{\rd \CW}{\rd A^i}\right),
}
}
where $\m$
is the  equivariant momentum map $\m: \CA \rightarrow Lie(\CG)^*$  
for the action of $\CG$ on $\CA$.
The potential energy $V$ in the sigma model above is given by 
\eqn\pah{
V =g_s^2\|D\|^2 + \sum_i \|H_i\|^2 
+\Fr{1}{g_s^2}\sum_i\bigl(\|\s^a\CL_a A^i\|^2 +\|\bar\s^a\CL_a A^i\|^2\bigr) 
+\Fr{1}{2g_s^2}\bigl\|[\s,\bar\s]\bigl\|^2.
}
Here for the auxiliary fields $H_i$ and $D$ the on-shell values 
\pag\ should be substituted. 
In the infrared limit $g_s\rightarrow 0$ the dominant contributions
to the path integral come from maps $A:\S\rightarrow \CM_\zeta$ to the 
locus of vanishing potential, modulo gauge transformations. We see from 
\pah\ that this is described by a symplectic quotient at level $\zeta$: 
\eqn\pae{
\CM_\zeta = \left(H_i^{-1}(0)\cap\m^{-1}(\zeta)\right)\!/\CG.
}
The group action preserves the condition $H_i=0$ and the subvariety 
$H_i^{-1}(0)\subset \CA$ inherits the complex and K\"{a}hler structures 
by restriction. The quotient space $\CM_\zeta$ 
inherits the K\"{a}hler structure from the ambient
space $\CA$ by the restrictions and the reduction. 
If $\zeta$ takes on a generic value, the group $\CG$ acts freely and
$\CM_\zeta$ is a smooth K\"{a}hler manifold. For such a case
the infrared limit of the theory can be identified with the 
non-linear sigma-model whose target space is $\CM_{\zeta}$. 
For non-generic $\zeta$ the quotient space develops singularities
or even may not exist at all. The infrared theory however should make 
sense also in these situations. For such cases however one always has 
some extra degrees of freedom not described by the moduli space, 
due to the extension of $\CA$ to $\CA_\CG$. If 
we vary $\zeta$ within the set of regular values the target space 
in general undergoes birational transformations. This is a physical 
realization of the variation of
symplectic quotients. The well-known relation between the symplectic
and geometrical invariant theory (GIT) quotients also is an important
part of the story \Kirwan. 
The essential point is that the condition $H_i=0$
is preserved by the complexified group action $\CG^\msbm{C}$, while
the condition $D=0$ is only preserved by the real group action. 
The complex gauge group in general does not act freely on 
the submanifold $H_i\inv(0)$, so that taking the quotient directly 
would lead to unwanted singularities. 
The GIT quotient considers the complex quotient by restricting 
to some {\it stable} subset $H_i^{-1}(0)_s\subset H_i^{-1}(0)$, on 
which the complexified gauge group acts freely, and sets 
$$
H_i^{-1}(0)\git\CG^\msbm{C} :=H_i^{-1}(0)_s /\CG^\msbm{C}.
$$
A proper condition for the stability should give rise to the
equivalence $H_i^{-1}(0)\git\CG^\msbm{C} =\CM_\zeta$
for generic and regular $\zeta$.

If $c_1(\CM_\zeta)=0$ the theory in the infrared limit is expected
to flow to a $\CN_{ws}=(2,2)$ superconformal theory. The chiral
operators of such a conformal theory by definition correspond 
to elements of $\CG$-equivariant de Rham or Dolbeault cohomology on 
the space $\CA$, carrying a suitable grading. 
The equivariant cohomology is a powerful mathematical tool. 
Note that if the moduli space is smooth, the equivariant 
cohomology on $\CA$ is equivalent to the ordinary cohomology 
on the moduli space. When the moduli space develops singularities, 
the ordinary cohomology is not well defined, while there is 
in general no problem with the equivariant cohomology 
on $\CA$. Thus we may even see the equivariant 
cohomology as a string-inspired generalization of ordinary 
cohomology. From the viewpoint of the gauged linear sigma model
this cohomology corresponds to the classical part of the story.
The quantum properties of the theory are even more
striking and beautiful, as exploited in many papers such 
as \GLSM\Wittengr\MorP.

\subsubsec{Some Finite Dimensional Examples}

We now consider some examples, mainly to indicate the comparison of our 
notation with the standard supersymmetry approach. 
We may also see this example as a sector of uncompactified 
matrix string theory, in the presence of extra branes. 
We start with $U(N)$ super-Yang-Mills in two dimensions. 
The theory contains a vector multiplet, containing the 
$U(N)$ connection one-forms (on the worldsheet) $v_{\pm\pm}$, 
the adjoint complex scalar $\s$ and fermions, as in \pad. 
The action functional of the theory is defined by the formula
$$
S(r,\bar r)= 
-\bs_+\bs_-\bbs_+\bbs_-\!\int_\S \!d^2\!x\,\tr \s\bar \s  
+ \Fr{r}{g_s} \bbs_+\bs_-\!\int \!d^2\!x\, \tr \s 
+\Fr{\bar r}{g_s} \bs_+\bbs_-\!\int \!d^2\!x\, \tr \bar\s.
$$
The expression is similar to the superspace expression, where we 
integrate over the fermionic coordinates, and replace the scalar $\s$ 
by the full vector superfield. This is equivalent, because the 
Berezin integral over the fermionic coordinates picks out exactly the 
supersymmetry transforms of the scalars, as in the expression above. 
We now generalize to the model for the Grassmannian considered in \Wittengr.
So we introduce $k$ complex scalars in the fundamental representation
of $\CG=U(N)$ and combine them into a $N\times k$ matrix $q$. In the space
of all such matrices we introduce a complex structure 
such that $\bbs_\pm q=0$. The $\CG$-action on such a space
is given by $q \rightarrow g q$ where $g\in \CG$. 
The above condition determines a chiral and an anti-chiral multiplet
and the supersymmetry transformation laws.
On the space of matrices $q$ we have a natural Hermitian 
structure given by $\int \!d^2\!x\, \tr q q^*$.
The corresponding action functional for this GLSM 
is then given by 
$$
S(r,\bar r)= 
\bs_+\bs_-\bbs_+\bbs_-\!\int_\S \!d^2\!x\,\tr(-\s\bar \s + qq^*)  
+ \Fr{r}{g_s} \bbs_+\bs_-\!\int \!d^2\!x\, \tr \s 
+ \Fr{\bar r}{g_s} \bs_+\bbs_-\!\int \!d^2\!x\, \tr \bar\s.
$$
The above action functional defines a GLSM for the 
Grassmannian $G(N,k)$ -- the space of $N$ complex planes in $\msbm{C}^k$.
After turning on the FI term $\zeta$  the model flows to a 
$\CN_{ws}=(2,2)$ non-linear sigma model whose target space
is $G(N,k)$, as can be seen from the localization equations \Wittengr.

\subsec{Digression: Comments on Gauged Non-Linear Sigma Models}

In this subsection we briefly comment on the non-linear
generalization of the gauged sigma-model and its
possible applications. 
The main motivation for this section is to compare
the proposal of Douglas \etal\ on matrix string
theory on Calabi-Yau \Douglas\DKO\ with our proposal.

To begin with we consider an example of an $U(1)$ theory.
Besides from the $\CN_{ws}=(2,2)$ vector multiplet
we introduce three complex scalars $\phi^i$, $i=1,2,3$, 
representing the holomorphic coordinates of a complex
$3$-dimensional K\"{a}hler manifold $X$, i.e. $\bbs_\pm \phi^i=0$. 
We assume that the $\phi^i$ are not charged under the $U(1)$ symmetry.
These conditions lead to three chiral multiplets and determine
the supersymmetry transformation laws. Let $K(\phi^i, \phi^{\bar\imath})$
be a K\"{a}hler potential for $X$.  Then the action functional
is defined by
\eqn\dna{
S = 
\bs_+\bs_-\bbs_+\bbs_-\!\int_\S \!d^2\!x\,\Bigl(-\s\bar \s 
+ K\bigl(\phi^i,\phi^{\bar\imath}\bigr)\Bigr)  
}
The resulting theory has two decoupled sectors; one is
the $U(1)$ SYM theory, where the $\phi^i$ are all zero, 
and the other is the supersymmetric non-linear sigma model.\foot{After 
adding the topological term defined by the pull-back of the K\"{a}hler
form on $M$, we have the standard action functional
for a $\CN_{ws}=(2,2)$ non-linear sigma model.}

The above model can easily be generalized to the non-abelian
version. We simply replace the gauge group by $U(N)$ and the 
complex scalars $\phi^i$ by $U(N)$ adjoint-valued
complex scalars, and consider the following action 
functional 
\eqn\dnb{
\eqalign{
S&= 
\bs_+\bs_-\bbs_+\bbs_-\!\int_\S \!d^2\!x\,\tr\Bigl(-\s\bar \s 
+ \BK\bigl(\phi^i,\phi^{\bar\imath}\bigr)\Bigr)  
\cr
&
+\Fr{1}{g_s}\bs_+\bs_- \!\int \!d^2\!x\,\tr \CW(\phi^i)
+\Fr{1}{g_s}\bbs_+\bbs_-\!\int \!d^2\!x\,\tr \overline\CW(\phi^{\bar\imath}),
}
}
where $\BK$ is a gauge covariant real functional and
$\CW(\phi^i)$ is a gauge covariant holomorphic functional
of the $\phi^i$. The above action functional has manifest
$\CN_{ws}=(2,2)$ supersymmetry. This type of model
has several interesting mathematical structures -- matrix
versions of K\"{a}hler metrics, Christoffel symbols, Riemann
tensor etc. However it is unclear what the conditions are 
for having a consistent quantum theory. 

As we assume that string theory is a consistent theory, 
at least the action for D-branes moving on curved space 
should be consistent. Furthermore, when this curved space is 
K\"ahler, we expect them to be described by a gauged non-linear 
sigma model as above, as argued by Douglas \Douglas. So the 
criteria for the gauged non-linear sigma model for describing 
D-branes on K\"ahler manifolds should then be sufficient. 
Such criteria, called the axioms of D-brane geometry, were formulated 
by Douglas, and were suggested to be used as a starting point 
for defining matrix theory on a curved space $X$ \Douglas. 
One of these axioms is the requirement that the moduli space 
of vanishing potential (modulo gauge symmetry) is the $N$th 
symmetric product of $X$. Comparing to our point of view, 
this moduli space should be identified with the quotient 
space \pae, so that we would need 
\eqn\dnc{
\left(H^{-1}_i(0)\cap \m^{-1}(0)\right)/U(N) = S^N X. 
}
Here $X$ is the base manifold represented by the center
of mass of the matrix coordinates $\phi^i$, i.e., 
$\{\Fr{1}{N}\tr \phi^i\}$. 
Another important axiom is the mass condition, which states that 
the off-diagonal matrix elements have masses proportional 
to the geodesic distance between the points on the diagonal. 
It is shown that the axioms require $M$ to be Ricci-flat, and fix
the holomorphic potential $\CW(\phi^i)$ to the
following minimal form
\eqn\dnd{
\CW(\phi^i) =\phi^1[\phi^2,\phi^3].
}
It is also shown that those axioms can be used to determine the
matrix version of a K\"{a}hler potential $\BK$ in terms of the 
K\"{a}hler potential of base manifold $X$ \DKO.

However it was demonstrated that such a model 
can be constructed only for Ricci-flat manifolds $X$ with 
vanishing six-dimensional Euler density \Wynter.\foot{An example
of such a manifold is the direct product $S\times C$ 
where $S$ is a hyper-K\"{a}hler surface.
Then we actually expect to have a $\CN_{ws}=(4,4)$ theory.
It is not even clear if a non-linear choice for $\BK$
always allows for a suitable $\CW(\phi^i)$ maintaining $\CN_{ws}=(4,4)$
supersymmetry. Clearly the choice \dnd\ is compatible only
with flat $S$.}
This result implies that matrix string theory compactifications 
on Calabi-Yau $3$-folds based on a $\CN_{ws}=(2,2)$
non-linear matrix sigma-model is not satisfactory so far.

In this paper we take an alternative approach. 
Instead of a $3N^2$ complex dimensional configuration space
(described by the matrices $\phi^i$) we consider the 
infinite dimensional linear space of all bundles on a 
Calabi-Yau $3$-fold. The infrared target space will then 
be the moduli space of stable bundles representing BPS 
configuration of D-branes on this Calabi-Yau. Also 
the gauge group will be infinite dimensional, 
consisting of gauge transformations in the bundles. 
Our model will be defined only on Calabi-Yau manifolds 
as we will see shortly. It would be conceivable 
that we could relate to a non-linear sigma model 
by integrating out massive degrees of freedom in the 
infrared theory. These massive modes would be 
higher modes on the Calabi-Yau where the bundles 
are defined. This would also reduce the gauge group 
to the finite dimensional $U(N)$ gauge group 
found in the approach of Douglas. In this way, 
the non-linear sigma model would turn up as an 
effective theory related to our linear sigma model.

\newsec{Sigma Model for Bundles on Calabi-Yau 3-Folds}

In this section we construct a $\CN_{ws}=(2,2)$ gauged linear 
sigma model whose target space is the infinite dimensional 
space of bundles on a Calabi-Yau $3$-fold.

\subsec{The Basic Settings}

We now come to the explicit construction of the model.
Consider a Calabi-Yau $3$-fold $X$ with K\"{a}hler form $\o$
and holomorphic $3$-form $\o^{3,0}$. 
We fix a rank $N$ $C^\infty$-bundle $E$ over 
$X$, endowed with a Hermitian structure. 
We fix the topological type of the bundle, by specifying its 
Chern character $\ch(E)$, or rather the Mukai vector 
$\ch(E)\sqrt{\widehat A(X)}$. For a Calabi-Yau $3$-fold, 
the Mukai vector is given by 
$$
Q = \biggl(\ch_0(E),\ch_1(E),
\ch_2(E)-\Fr{p_1(X)}{48}\ch_0(E),
\ch_3(E)-\Fr{p_1(X)}{48}\ch_1(E)\biggr),
$$ 
where $p_1(X)$ is the first Pontryagin class of the 
Calabi-Yau manifold. 
We may sum over different topological types later.
The bundles may be seen as describing D-branes wrapped 
around the Calabi-Yau manifold $X$. The D-brane charges are 
precisely given by the components of the Mukai vector 
\HarveyMoore\Inflow.
We will denote these D-branes by their part wrapped 
around the Calabi-Yau. For example, the rank $N=\ch_0(E)$ 
corresponds to the number of $D_6$-branes wrapped around $X$ and 
more generally the charges $Q_{3-n}(E)\sim\ch_{3-n}(E)$ 
correspond to $D_{2n}$-branes wrapped around cycles in $X$ \Vafa. 
In a type IIB setting, these branes do not exist in the total 
ten-dimensional space-time. To get a type IIB brane, one should 
wrap the brane around another direction; this will of course be
 the spatial direction of the matrix string. The $D_6$-brane 
(in our notation) then corresponds in the full type IIB 
string theory to a $D7$-brane. 

We denote by $Lie(G)$ the Lie algebra of $G=U(N)$ and by 
$\End(E)=E\otimes E^*$ the bundle of endomorphisms.
Let $\CA$ be the infinite dimensional space of all connections, and 
$\CG$ the infinite dimensional group of gauge transformations 
$g: X\rightarrow G$. As usual $\CA$ is an affine space, 
and a tangent vector is represented by $\d A\in\O^1(X,\End(E))$. 
We want to use this (infinite dimensional) linear space $\CA$ and the 
group $\CG$ as the target space and gauge group respectively 
for a GLSM. 
To fit the above data in the framework described in the previous section we
need some preparations -- complex structure, K\"{a}hler potential,
Dolbeault equivariant cohomology and a holomorphic potential leading
to integrability.

Given the complex structure on $X$, we may introduce a complex structure 
on the space of connections $\CA$ as follows. 
Let $A$ denote a connection one-form, which is decomposed into
its holomorphic and antiholomorphic components $A=A^{1,0} + A^{0,1}$. 
One introduces a complex structure 
$\CA$ by declaring $\d A^{0,1}\in\O^{0,1}(X, \End(E))$ to be a holomorphic
tangent vector. Endowed with this complex structure $\CA$ becomes an 
infinite dimensional flat K\"{a}hler manifold with K\"{a}hler form 
$\Bz$ given by 
\eqn\baa{
\Bz(\d A^{1,0},\d A^{0,1}) =  
\Fr{i}{8\pi^2}\int_X \tr (\d A^{1,0}\wedge\d A^{0,1})\wedge\o\wedge\o.
}
The group of gauge transformations $\CG$ acts with isometries on 
this space. 
The K\"{a}hler potential for $\Bz$ is given by
\eqn\bab{
\Fr{1}{4\pi^2}\CK\bigl(A^{1,0},A^{0,1}\bigr) =
\Fr{i}{8\pi^2}\int_X \k\, \tr (F\wedge F)\wedge\o,
} 
where $\k$ is a K\"{a}hler potential for $\o$.
Thus both the complex structure and K\"{a}hler moduli in 
$\CA$ depend on those in the base space $X$.

Now  we consider $\CG$-equivariant differentials $\bs$ and $\bbs$
on $\CA$ (they constitute the operators $\rd_\CG$ and 
$\bar\rd_\CG$ in \pab)
such that
\eqn\bad{
\eqalign{
\bs A^{0,1} &=i\p^{0,1},\cr
\bbs A^{0,1}&=0,\cr
\bs A^{1,0}&=0,\cr
\bbs A^{1,0}&=i\bar\p^{1,0},\cr
}\qquad
\eqalign{
\bs \p^{0,1}&=0,\cr
\bbs\p^{0,1} &= -\Dpp \phi,\cr
\bs\bar\p^{1,0}&=-\Dp \phi,\cr
\bbs\bar\p^{1,0}&=0,\cr
}
\qquad
\eqalign{
\bs\phi=0,\cr
\bbs\phi=0,\cr
}
}
where $\p^{0,1}\in \O^{0,1}(X, \End(E))$ represents a holomorphic
(co)-tangent vector on $\CA$ and the adjoint scalar 
$\phi\in \O^0(X, \End(E))$
is the generator of an infinitesimal $\CG$-action on $\CA$.
We have $\{\bs,\bbs\}A = -id_{\!A}\phi$ satisfying \pab.
Using these equivariant differentials, we have an equivariant 
K\"{a}hler identity 
\eqn\bae{
\eqalign{
\widetilde\Bz&=
\Fr{i}{4\pi^2}\bs\bbs \CK\bigl(A^{1,0},A^{0,1}\bigr)\cr
&= \Fr{i}{4\pi^2}\int_X \tr (i\phi F)\wedge\o\wedge\o +\Fr{i}{4\pi^2}
\int_X \tr\bigl(\p^{0,1}\wedge\bar\p^{1,0}\bigr)\wedge\o\wedge\o,
}
}
where the second term can be identified with the K\"{a}hler form $\Bz$
and the first term is the moment map $\phi^a\m_a$,
$\m:\CA\rightarrow Lie(\CG)^*=\O^{6}(X,\End(E))$ for the action 
of $\CG$ on $\CA$,
\eqn\baf{
\m(A) =  \Fr{1}{4\pi^2} F\wedge\o\wedge\o
= \Fr{1}{12\pi^2}(\L F) \,\o\wedge\o\wedge\o,
}
where $\L$ is the adjoint of wedge multiplication by $\o$.
$\widetilde\Bz$ is known as an equivariant K\"ahler form.

The K\"ahler structure on the space of bundles does not give enough 
structure for our purpose. The moduli space of bundles, which in 
the end will be identified with the infrared target space of our  
model, should be a finite dimensional space. The space of gauge 
equivalence classes of bundles however can never be finite dimensional.
This can easily be seen as follows. 
Using the K\"ahler structure on $X$, we can decompose the curvature 
two-form of the bundle $E$ into type according to 
$F = F^{2,0} + F^{1,1} + F^{0,2}$. Using the moment map $\m$, 
we can restrict only the $F^{1,1}$ part of the curvature. The 
$F^{0,2}$ part however will not be restricted, thus leading 
to an infinite dimensional space of deformations. There is 
a natural way to further restrict the set of gauge bundles. 
To this end, we consider the infinite dimensional subvariety 
$\CA^{1,1}$ of all connections for which the curvature is of 
type $(1,1)$, so 
\eqn\bac{
F^{0,2}_A=0. 
}
Thus $\Dpp^2=0$ for $A\in\CA^{1,1}$. This condition endows the 
bundle $E$ with a holomorphic structure. 
The moduli space of holomorphic bundles is the set of 
bundle isomorphism classes. It can be given by the following
complex quotient 
\eqn\bac{
\CA^{1,1}/\CG^\msbm{C}, 
}
where $\CG^{\msbm{C}}$ is the complexification of $\CG$.
As discussed in the last section, we can restrict the model 
by adding a holomorphic potential $\CW$ as in \paf\ to 
the model. Indeed, we are even forced to do so, as the 
moduli space \pae\ would otherwise not be finite dimensional. 
The holomorphic potential should be a holomorphic functional of 
the coordinates $A^{0,1}$ on $\CA$. Furthermore, it should be 
gauge invariant at least for gauge transformations connected 
to the identity. This essentially fixes the holomorphic 
potential to the holomorphic Chern-Simons functional, 
which is given by
\eqn\bah{
\CW(A^{0,1}) = \int_X \o^{3,0}\wedge\tr
 \Bigl(\Fr{1}{2}A^{0,1}\wedge\bar\rd A^{0,1}
+ \Fr{1}{3}A^{0,1}\wedge A^{0,1}\wedge A^{0,1}\Bigr).
}
Note that this potential through \pag\ gives rise to exactly 
the condition \bac\ in the infrared, 
\eqn\bah{
\Fr{\d \CW}{\d A^{0,1}}=0\longrightarrow F^{0,2}_A=0.
}

Note that the construction of this holomorphic potential 
is only possible on a Calabi-Yau manifold, 
as it makes use of the holomorphic 3-form $\o^{3,0}$. 
This functional was first considered by Witten \Wittencss, 
but was interpreted there as the action functional rather
than a superpotential for his effective open string theory. 
We come back to the relation of his model to ours later 
in this paper.

We have now defined all the data needed for the 
construction of a $\CN_{ws}=(2,2)$ gauged linear
sigma model associated with the infinite 
dimensional pair $(\CA,\CG)$.

\subsec{The $\CN_{ws}=(2,2)$ GLSM}

To construct the GLSM explicitly, we consider a vector bundle 
$\tilde E$ over $X\times\S$, with structure group $U(N)$.
The group of all gauge transformations in this vector 
bundle will be denoted $\tilde\CG$, and 
$\tilde \CA$ is the space of all connections on $\tilde E$.
We denote by $E$, $\CG$ and $\CA$ the restrictions of $\tilde E$, 
$\tilde \CG$ and $\tilde\CA$ respectively to $X\times\{pt\}$.
We will use spinor notation for the world-sheet $\S$ 
and differential form notation for the Calabi-Yau $X$.  
The supercharges evaluated at a point $\{pt\}\in \S$ are differentials
of balanced $\CG$-equivariant Dolbeault cohomology on
$\CA$. The $\CN_{ws}=(2,2)$ supersymmetry algebra is 
defined by the commutation relations \aaa.
In terms of infinitesimal generator $\ep_-,~\bar\ep_-$, 
which are sections of $K^{-1/2}_\S$, and 
$\ep_+,~\bar\ep_+$, which are sections of $\overline K^{-1/2}_\S$, 
we denote $\d = \bar\ep_- \bs_+ +\bar\ep_+\bs_- + \ep_+\bbs_- +\ep_+\bbs_+$.
The supercharges transform as scalars on the Calabi-Yau $3$-fold $X$.

The GLSM related to gauge bundles on the CY has the following field 
content.
First, there is the $\CN_{ws}=(2,2)$ vector multiplet
as in the diagram \pad.  The world-sheet vector multiplet 
transforms as an adjoint valued scalar on $X$.
The explicit transformation laws are given in Appendix A. 
The covariant derivatives on the worldsheet are defined  
as $\nabla_{\!\pm\pm}=\rd_{\pm\pm}+v_{\pm\pm}$, and its 
curvature is $f_\S=[\nabla_{\!++},\nabla_{\!--}]$.
The fields in the vector multiplet $(v_{\pm\pm},\eta_\pm,D,\s)$
will have worldsheet scaling dimensions $(1,3/2, 2,0)$.
Secondly we have chiral and anti-chiral multiplets
from the connection one-form $A=A^{1,0}+A^{0,1}$ on $X$
According to our choice of complex structure on $\CA$,
we build up chiral $(\bbs_\pm A^{0,1}=0)$ multiplets 
from $A^{0,1}$. The transformations in the chiral multiplet 
are as in the following diagram
\eqn\bag{

\matrix{
\p^{0,1}_- &\mapl & A^{0,1} & \mapr & \p^{0,1}_+ \cr
  & \rlap{\lower.3ex\hbox{$\scriptstyle \bs_{\!+}$}}\searrow
 && \swarrow\!\!\!\rlap{\lower.3ex\hbox{$\scriptstyle \bs_{\!-}$}} & \cr
&& H^{0,1} &&
}
}
The explicit transformation rules for a chiral multiplet 
can be found in Appendix A. 

Similarly, the fields in the $A^{1,0}$ multiplet form 
anti-chiral ($\bs_\pm A^{1,0}=0$) multiplets
\eqn\bag{

\matrix{
\bar\p^{1,0}_- &\mapbl & A^{1,0} & \mapbr & \bar\p^{1,0}_+ \cr
  & \rlap{\lower.3ex\hbox{$\scriptstyle \bbs_{\!+}$}}\searrow
 && \swarrow\!\!\!\rlap{\lower.3ex\hbox{$\scriptstyle \bbs_{\!-}$}} & \cr
&& H^{1,0} &&
}
}

Note that, as $A$ is a gauge field, any commutator with $A$ 
should be replaced by a covariant derivative. We will write 
the covariant Dolbeault operator related to $A^{0,1}$ and 
$A^{1,0}$ as $\Dpp=\bar\rd+A^{0,1}$ and $\Dp=\rd+A^{1,0}$ 
respectively.

The  left and right $U(1)$ $\CR$-charges $(J_L,J_R)$ 
for the chiral and anti-chiral matter multiplets are given in Table 2. 

\lin{Table 2}
$$

\matrix{
    & A & \bar\p_+^{1,0} & \bar\p_-^{1,0} & \p_+^{0,1} & \p^{0,1}_- & H^{1,0} & H^{0,1}
\cr
J_L & 0 & -1             & 0              & +1         & 0          & -1      & +1
\cr
J_R & 0 & 0              & -1             & 0          & +1         & -1      & +1  
}
$$

The action functional can be given as in the general 
formula \paf. Using the particular form of the K\"ahler 
and super potential as given above, this can be written 
\eqn\action{
\eqalign{
S(r,\bar r)= 
& - 
\bs_+\bs_-\bbs_+\bbs_-\!\int_\S \!d^2\!x\!
\int_X\! d\m_X \tr( \s\bar\s)
+\bs_+\bs_-\bbs_+\bbs_-\!\int_\S \!d^2\!x\,
\CK(A^{1,0}, A^{0,1})
\cr
&
+ \Fr{r}{g_s} \bbs_+\bs_-\!
\int_\S \!d^2\!x\!\int_X d\m_X\tr(\s) 
+c.c.
\cr
&
+\Fr{1}{g_s}\bs_+\bs_- \!\int_\S \!d^2\!x\, \CW\bigl(A^{0,1})
+c.c.,
}
}
where $d\m_X$ denotes the volume form on $X$.
Since $\bs_+\bs_-$ has $U(1)$ $\CR$-charge $(1,-1)$, $\CW(A^{0,1})$ 
should have charges $(-1,1)$ to preserve the $\CR$-symmetry.
Since $A$ naturally has charges $(0,0)$, the 
only choice left is to assign charges $(-1,1)$ to $\o^{3,0}$.

The model is  realized
as an eight-dimensional $U(N)$ gauge theory on the product manifold
$\S\times X$. Some of the supercharges in this theory are broken due to 
the nontrivial background. The surviving supercharges should be 
covariantly constant on $X$, while spinors on the worldsheet $\S$. They 
can then be identified with scalars on $X$, by a trivial twist.\foot{
The model can thus be identified with a twisted version of eight-dimensional 
super Yang-Mills, similar to the approach of \BJSV\ in four dimensions.}
These are our supercharges $\bs_\pm,\bbs_\pm$. 
As explained, we regard our model as a linear sigma model in two dimensions 
with infinite dimensional target space $\CA$ and gauged 
isometry group $\CG$. 
We may regard the Calabi-Yau $3$-fold as a parameter space
describing a continuous family of pairs $(\CA,\CG)$. As we discussed
earlier $\CA$ inherited both its complex and K\"{a}hler structure 
from $X$. 
There is no inconsistency here since the supercharges
are topological when restricted to $X$. The path integral is then 
independent of the size of $X$ and we can take the limit 
$vol(X)\rightarrow 0$ to recover the two-dimensional sigma-model on $\S$. 
The number of bose and fermi fields coincides with those
of $\CN=1$ SYM theory in ten dimensions.

The six-dimensional model was also considered in \Baulieu\ in terms 
of a $\CN_T=1$ cohomological field theory, as a special example of a 
more general construction of cohomological theories for moduli spaces 
of bundles.

\subsec{The Infrared Limit}

For finite string coupling constant $g_s$ arbitrary bundles
on the Calabi-Yau contribute to the path integral.
In the infrared limit $g_s\rightarrow 0$ the dominant contributions
to the path integral come from the space of all maps from the 
worldsheet $\S$ to the vanishing locus of the potential $V$ 
(given in \pah) modulo the $\CG$-action.
With our choice of K\"ahler and super potential, these are 
determined by the following conditions 
\eqn\bba{
\eqalign{
F^{0,2}=0,\cr
\L F -\zeta I=0,
}
}
and
\eqn\bbab{
\eqalign{
d_{\!A}\s=0,\cr
[\s,\bar\s]=0.
}
}
The connections solving the first two equations are called
{\it Einstein-Hermitian} (EH) connections \Kobayashi. 
They correspond to Einstein-Hermitian vector bundles. 
The moduli space of EH connections is the symplectic quotient
\eqn\bbc{
\CM_{EH} = (\CA^{1,1} \cap \m^{-1}(\zeta))/\CG.
}
We denote by $\CM^*_{EH}$ the moduli space of irreducible
EH connections. By the Donaldson-Uhlenbeck-Yau theorem
the moduli space $\CM_{EH}^*$ is diffeomorphic to
the moduli space of $\o$-stable holomorphic vector bundles
defined by the GIT quotient \Donaldson\UY
\eqn\bbd{
\CA^{1,1}\git\CG^\msbm{C}=\CA^{1,1}_s/\CG^\msbm{C}
}
If the connection is irreducible, the condition $d_{\! A}\s=0$ 
implies that $\s=0$. The two equations in \bbab\ have non-trivial
solutions if an EH connection is reducible. Typically a reducible
connection gives rise to a singularity in $\CM_{EH}$. The equations 
\bba\ do not guarantee that we always have irreducible connections. 
From the equations in \bbab\ we see that such a 
reducible connection also gives rise to a non-compact direction in 
the localization manifold. These non-compact directions are not 
specially related to these singularities; the moduli space $\CM_{EH}(X)$ 
is non-compact even if there are no reducible connections.\foot{
According to Gieseker the moduli space of stable bundles is an 
open subset of the moduli space of semi-stable bundle. This provides
the Gieseker compactification for the moduli space of stable bundles
by taking the closure. The definition of stable bundles involves
torsion free sheaves. One may consult a nice book \Kobayashi\
for details.
}

A fundamental result of Witten for $\CN_{ws}=(2,2)$ gauged
linear sigma models states that the physics of the infrared super
conformal theory is smooth even if the target space develops
singularities \GLSM. In many respects 
the string theory compactifies the target space and we may
constructively identify the infrared target space as 
the moduli space of semistable torsion free sheaves on $X$.
Note that the notion of (semi-)stability is variable depending
on the polarization. If one changes the polarization the moduli space
may undergo a sequence of birational transformations.
Witten's analysis implies that the physics is independent
of the polarization. In our case the K\"{a}hler form $\o$ 
on $X$ determines the polarization of stability ($\o$-stability). 
The (semi-)stability plays an important role in the model, and is 
also related to stability of bound states of wrapped D-branes 
\HarveyMoore. 

We briefly discuss the role of the FI term $\zeta$. In general, 
the second equation in \bba\ is called the weak Einstein condition. 
The parameter $\zeta$ is constant along the worldsheet $\S$. 
However, $\zeta$ can be a real function on $X$. If it is constant 
the second equation in \bba\ is called the Einstein condition with 
factor $\zeta$. In the more general case one can relate the weak 
Einstein condition to the Einstein condition by a conformal change 
of the Hermitian metric on the bundle $E$. We will here take 
$\zeta$ to be constant. The Einstein condition then directly 
implies that $\zeta$ is given by
\eqn\gaa{
\zeta =\left(\int_X \ch_1(E)\wedge\o\wedge\o\right)\Big/
\left(\Fr{N}{6\pi}\int_X \o\wedge\o\wedge\o\right)
}
thus depends only the the cohomology classes of $\o$ and $c_1(E)$.

We may now conclude that
our model flows to a non-linear sigma model for a 
Calabi-Yau with semi-stable bundles. 
We expect that the resulting sigma-model is superconformal since 
$\CM_{EH}$ inherits a Calabi-Yau structure.
The case of rank $N$ corresponds to a Calabi-Yau $3$-fold
with $N$ $D_6$-branes bounded with 
$D_{2n}$-branes classified by the Chern characters $\ch_{3-n}(E)$. 
For example the equation \gaa\ implies that if the volume of the 
$D_{4}$-brane collapses to zero we should have $\zeta=0$.
The condition to preserve supersymmetry translates to
stability. EH bundles can only exist when the following 
topological condition is met 
\eqn\bbd{
\int_X \bigl(2N \ch_2(E) - \ch_1(E)\wedge \ch_1(E)\bigr)\wedge\o \geq 0,
}
where the equality holds if and only if $E$ is projectively flat. 
If we do not have any $D_2$- and $D_4$-branes the bundles are flat.  
For $\ch_1(E)=0$, that is when there are no $D_4$-branes, the above 
condition reduces to
\eqn\bbdx{
\int_X \ch_2(E)\wedge\o \geq 0.
}
This is a direct generalization of the condition in four dimensions 
that only ASD connections survive. The more general condition \bbd\ 
is just a slight modification of this restriction.

\subsec{Reduction to Matrix String Theory of Five-Branes Compactified
on a $K3$ Surface}

Here we briefly comment on relation with matrix string theory of
five-branes, whose world-volume is compactified on a $K3$ surface.

We consider the case that that the Calabi-Yau $3$-fold $X$ is
a product manifold $X=K3\times T^2$. We will consider the limit of 
vanishing $T^2$. Then we can $T$-dualize along the $T^2$-direction to 
reduce our model to a gauged linear sigma model for bundles on $K3$.
This amounts to the simple dimensional reduction along the 
$T^2$ direction. The vector $Q$ of D-brane charges reduces to 
the Mukai vector for the bundle on $K3$. 
The connection $(0,1)$-form in six dimensions then decomposes
into $A^{0,1}\oplus \t$, where $A^{0,1}$ is the component along 
the $K3$ and $\t$ the component along the torus, which becomes a 
complex adjoint scalar on the $K3$ surface. More generally, the 
chiral multiplet \bag\ decomposes into two chiral multiplets; 
one including the connection $(0,1)$-form on the $K3$ surface, 
\eqn\cag{

\matrix{
\p^{0,1}_- &\mapl & A^{0,1} & \mapr & \p^{0,1}_ + \cr
 & \rlap{\lower.3ex\hbox{$\scriptstyle \bs_{\!+}$}}\searrow
 && \swarrow\!\!\!\rlap{\lower.3ex\hbox{$\scriptstyle \bs_{\!-}$}} & \cr
&& H^{0,1} &&
},
}
and the other with the adjoint complex scalar $\t$, 
\eqn\cagk{

\matrix{
\l_- &\mapl & \t & \mapr & \l_+ \cr
 & \rlap{\lower.3ex\hbox{$\scriptstyle \bs_{\!+}$}}\searrow
 && \swarrow\!\!\!\rlap{\lower.3ex\hbox{$\scriptstyle \bs_{\!-}$}} & \cr
&& H &&
}.
}

After the above reduction the holomorphic superpotential 
$\CW$ in \bah\ reduces to
\eqn\cvh{
\CW_4 =\int_{K3}\o^{2,0}\wedge \tr\t F^{0,2},
}
where $\o^{2,0}$ denotes the holomorphic symplectic 
form on the $K3$ surface.
Since $\o^{2,0}$ is a nowhere vanishing non-degenerated
$2$-form we may regard $\t$ as a holomorphic two-form 
$\t^{2,0}:=\t\o^{2,0}$. This should of course be extended to 
the full chiral multiplet \cagk. 
Similarly the K\"{a}hler potential $\CK$ in \bab\
decomposes into
\eqn\cvi{
\CK_4 = \int_{K3} \tr\Bigl(i\k F\wedge F 
 - \t^{2,0}\wedge\bar\t^{0,2}\Bigr),
}
where $\k$ is a K\"{a}hler potential for the $K3$.
The action functional is given by the same formula \action, where
$\CW$ and $\CK$ are replaced by their respective expressions
given above.

The worldsheet supersymmetry of the resulting model 
enhances to $\CN_{ws}=(4,4)$ supersymmetry. The adjoint
chiral multiplet with bosonic component $\t$ in \cagk\
combines with the $\CN_{ws}=(2,2)$ vector multiplet in \pad\ 
into a $\CN_{ws}=(4,4)$ vector multiplet. 
In this correspondence, the scalars $\t$ and $\s$ combine 
into a self-dual 2-form $B^+$ and a real scalar $C$, as follows 
\eqn\cvk{
B^+ = \t\o^{2,0}+\bar\t\o^{0,2}+\Im\s\,\o,\qquad  
C=\Re \s.
}
This gives exactly the field content of the twisted $\CN=4$ 
SYM on $K3$ studied by Vafa and Witten \VW. 

In the infrared limit the theory reduces to a $\CN_{ws}=(4,4)$
non-linear sigma model. The target space is given by
the solutions of the following equations, modulo the
gauge transformations
\eqn\bbal{
\eqalign{
F^{0,2}=0,\cr
\L F -\zeta I=0,
}
}
and 
\eqn\bbam{
\eqalign{
d_{\!A}\s=d_{\!A}\t=0,\cr
[\s,\bar\s]=[\t,\bar \t]=[\s,\t]=[\bar\s,\t]=0.
}
}
Note that the EH condition reduces to the condition of 
ASD connections on $K3$.\foot{This is deformed by the 
FI parameter $\z$, which also is a natural deformation 
in the $K3$ situation \Robbert.} If the EH connection is 
irreducible, the equations can only be solved by $\s=\t=0$.
Then the target space of this model is the 
moduli space of stable bundles on $K3$, which can be identified with
the moduli space $\CM^*_{ASD}$ of irreducible anti-self-dual 
(ASD) connections on the $K3$ surface. 
Our  model in the infrared limit can be identified 
with the matrix string theory of the five-brane compactified 
on $K3$ discussed in 
\mbh\DVVrev\Robbert, which was based on 
orbifold conformal field theory.\foot{Our 
gauge theoretic description has an obvious
problem due to the non-compactness of the moduli space of
instantons. We better regard the infrared target space as
the moduli space of torsion-free coherent sheaves, 
as emphasized in \HarveyMoore. If the moduli space contains 
strictly semi-stable sheaves, which is inevitable in certain 
cases, the identification with orbifold conformal theory may 
be problematic. The torsion free sheaves are also relevant to 
matrix string theory in the presence of $k$ five-branes \ABS. 
The infrared target space is then the moduli space of 
torsion-free sheaves on $\R^4$ via the ADHM description.}
Our reduced model describes the matrix string theory of the five-brane 
as a gauged linear sigma model in accordance with our general
philosophy.\foot{The $\CN_{ws}=(4,4)$ world-sheet supersymmetry
evaluated at a point on the worldsheet defines a balanced 
$\CG$-equivariant hyper-K\"{a}hler cohomology \HP.}
This moduli space is known to be birational to a 
symmetric symmetric product of a (dual) $K3$ surface   
$\CM^*_{ASD}=S^N\widetilde{K3}$ in general. 
In this way, we can identify the infrared limit of the model on $K3$ 
as a system of (weakly coupled) fundamental strings on 
$\widetilde{K3}$, in accordance also with the axioms of 
D-brane geometry \Douglas. 
The model in the infrared limit describes the Higgs branch
of a $D1-D5$ system where the $D5$-brane is wrapped around $K3$. 
By applying T and S duality the model describes the matrix
string theory of the five-brane wrapped around $K3$.
The model describes six-dimensional interacting micro matrix
strings and can be regarded as a microscopic definition of
IIB string theory on $AdS_3\times S^3\times K3$ due to a
celebrated conjecture of Maldacena \JuanLN.

The above identification is evidence that
our model for bundles on Calabi-Yau can be 
regarded as the matrix string theory of Calabi-Yau compactifications.
The  model was already suggested in \DPS\ based on a similar
approach. If we perform dimensional reduction along the world-sheet
our reduced model becomes the Vafa-Witten model of $\CN=4$
super-Yang-Mills theory on $K3$. The partition function
of this topological field theory computes the Euler 
characteristic of the moduli space of instantons \VW.

\subsec{Decoupling the Bulk Degrees of Freedom}

Stable bundles appear naturally in the context of non-perturbative
string theory \HarveyMoore. They correspond to the stable
BPS configurations of branes wrapped around non-trivial
cycles in the compactified part $X$ of the bulk space time $Z$.
These are also naturally associated with extremal black-hole
solutions of the low energy effective supergravity.
A suitable counting of the number of stable orbits corresponds
to counting the microscopic degrees of freedom of these black 
holes. The asymptotic growth of the degeneracy then gives 
the black-hole entropy. In our context the
natural object to study is the elliptic genus directly
relevant to the four-dimensional black-hole.
The semi-stable orbits which are not stable correspond to
marginally stable brane configuration. They correspond to branes
wrapped around certain vanishing cycles in $X$. Physically 
such states are new massless (tensionless) states free to 
escape to the bulk $Z$. Indeed, in the strictly semi-stable 
case the equations \bbab\ (or \bbam\ in the case of $K3$) 
allow for nontrivial solutions for $\s$ (and $\t$), 
which describe the degrees of freedom outside the space $X$. 
Such an orbit also introduces singularities 
in the moduli space, indicating that the degrees of 
freedom of the bundle do not contain all the information 
necessary to describe the system.

Now we examine the above properties in the context of our models
for $X=CY_3,~K3$. We note that the equations for the infrared 
target space \bba\bbab\ or \bbal\bbam\ are precisely the equations 
for BPS configurations for D-branes wrapped around $X$ \HarveyMoore.\foot{
As was remarked in \HarveyMoore, this BPS condition should be valid 
only in the limit of vanishing string coupling. This is consistent 
with our description, as we find these equations only at the infrared limit.} 
Formally, from the viewpoint of the string world-sheet, the infrared limit
corresponds to the limit where the bulk string coupling constant
becomes zero. The string theory then flows to a superconformal
non-linear sigma-model whose target space is the moduli space of 
semi-stable bundles together with the linear space spanned by the 
zero-modes of the equations in \bbab\ or \bbam\ for the adjoint 
complex scalars ($\s$ for $CY_3$ and $(\s,\t)$ for $K3$). 
These zero-modes represent the bulk degrees of freedom transverse
to the compact space $X$. When the brane
configuration is stable there are no zero-modes for the adjoint scalars. 
The stable bundles hence represent configurations of branes which 
are completely decoupled from the bulk. The matrix string only 
propagates on the compact space $X$.

Consequently the infrared superconformal theory on the string world-sheet
involving stable bundles describes the decoupled matrix string
theory. The M(atrix) conjecture as well as Maldacena's conjectures state
that such a theory is dual to string/M theory in a non-trivial 
background given by the near horizon limit \Why\JuanLN.
As far as the description in terms of matrix string theory
is concerned the decoupling mechanism is exactly the
same for both $CY_3$ and $K3$. Thus it seems to be natural
to conjecture that the infrared conformal theory for the $CY_3$ case
has an analogous dual description. The natural conjecture
is duality with IIB string theory on $AdS_3\times S^1\times CY_3$. 
Here the $AdS_3$ space comes from the worldsheet and the norm of 
$\s$, while the $S^1$ is described by the phase of $\s$. 
There are several problems with such a relation.  
First of all, compactification on a Calabi-Yau $3$-fold 
needs, in terms of type IIB, 7-branes wrapped around the Calabi-Yau. 
These are hard to describe, especially in the context of M-theory. 
Also, the near-horizon geometry for these 7-branes is not so well 
behaved. 
Secondly, the dilaton in this case is not constant, so that we 
can not tune it to a small value. This implies that we can not 
identify a region in moduli space where the string is 
weakly coupled. This makes it very hard indeed to 
make use of such a correspondence. 

As a superconformal non-linear sigma-model the chiral rings can be described
by a topological sigma-model \chiralrings\TSM\Wittenmirror.
These topological quantities will be important ingredients for checks 
of the M(atrix) and Maldacena conjecture. Another interesting
quantity is the elliptic genus (the half-twisted model).
For the $K3$ case the elliptic genus of the world-sheet superconformal
theory \orbifolds\ is used to test the duality \deBor.

\newsec{Applications: Twisted Models}

Given a $\CN_{ws}=(2,2)$ GLSM natural objects to study 
are supersymmetric indices -- the Euler characteristic,
the elliptic genus, and the chiral rings.
Those topological and pseudo topological quantities
contain interesting information both for physics and mathematics.
The (pseudo) topological quantities are most naturally 
studied using topologically twisted versions of the 
supersymmetric theory we have been studying. These twisted 
versions are the subject of this section. It could 
also be a starting point for the study of the generalized
mirror conjecture \Vafa\ from a sigma model viewpoint 
\Wittenmirror\kstheory\GLSM.
An obvious benefit of the GLSM is that those (pseudo) topological
quantities attributed to the infrared superconformal non-linear
model can be evaluated in a different regime of the
theory.

The Euler characteristic of the moduli space of stable EH 
bundles corresponds to the holomorphic Casson invariant 
which was defined by Thomas \Thomas\ThomasTh. The elliptic
genus is the stringy generalization of this quantity. The 
elliptic genus is particularly relevant for the 
four-dimensional black-hole entropy. 
The correlation functions of the $A$-model correspond to the 
quantum intersection pairing of the moduli space of stable bundles. 
This gives a stringy generalization of Donaldson-Witten type
invariants on a Calabi-Yau $3$-fold. We also remark
that the mathematical definition of these classical
invariants involve hard technical obstacles.
As a folk theorem one expects that string theory
may soften many, if not all, of these problems.

We begin with the description of the $A$-model and then 
proceed with the $B$-model. The half-twisted
model computing the elliptic genus can be treated along
the lines of the $A$-model. We will not consider it here. As usual 
we perform a Wick-rotation on the worldsheet, and use holomorphic 
coordinates on $\S$.

\subsec{The $A$-Model}

The $A$-model (and the half twisted model) can be defined following the 
standard recipe \Wittenmirror\GLSM. The observables of the theory are given
by $\CG$-equivariant differential forms on the target space $\CA$. 
In the infrared limit these observables can be identified with differential 
forms on the moduli space $\CM_{EH}$, and therefore flow to the usual 
observables in a topological non-linear sigma model \TSM.

In the $A$-model the twist on the worldsheet is performed such that 
$\ep_+=\ep$ and $\bar\ep_-=\bar\ep$ become worldsheet scalars. They 
are then set equal to constants on $\S$. The other generators 
$\bar\ep_+$ and $\ep_-$ are set to zero. Thus we are keeping 
the supercharges $\bs_+$ and $\bbs_-$, which 
now transform as world-sheet scalar under the two-dimensional
rotation group. As there is no source for confusion, we leave out 
the subscript $\pm$ in the rest of this subsection. 
The BRST operator of the model is then given by 
$\d=\bar\ep\bs +\ep\bbs$. 
The resulting model computes the quantum cohomology
ring of the moduli space of holomorphic vector bundles over
the Calabi-Yau $3$-fold $X$. 
The twisting maps the $\CN_{ws}=(2,2)$ vector multiplet to a basic 
vector multiplet and an anti-ghost multiplet according to 
$$
\eqalign{
(v_{++},\bar\eta_+)\rightarrow & (v_{z},\th_z),\cr
(v_{--},\eta_-)\rightarrow & (v_{\bar z},\th_{\bar z}),\cr
(\bar\s,\eta_+,\bar\eta_-,D) \rightarrow & (\bar\s,\eta,\bar\eta,D).
}
$$
The (anti) chiral multiplets containing the target space 
vector fields are twisted in the following way, giving rise to basic 
multiplets and anti-ghosts 
$$
\eqalign{
(A^{1,0},\bar\p^{1,0}_-) \rightarrow& (A^{1,0},\bar\p^{1,0}),\cr
(A^{0,1},\p^{0,1}_+) \rightarrow& (A^{0,1},\p^{0,1}),
}\qquad
\eqalign{
(\bar\p^{1,0}_+,H^{1,0}) \rightarrow& (\bar\c^{1,0}_z,H^{1,0}_z),\cr
(\p^{0,1}_-,H^{0,1}) \rightarrow& (\c^{0,1}_{\bar z},H^{0,1}_{\bar z}). 
}
$$

The BRST transformation laws for the basic fields are
\eqn\ujb{
\eqalign{
\d A^{1,0} 
	=& i\ep\bar\psi^{1,0},
\cr
\d A^{0,1} 
	=& i\bar\ep\psi^{0,1} ,
\cr
\d v_{z} =& i\ep\th_{z},
\cr
\d v_{\bar z} =& i\bar\ep\th_{\bar z},
}\qquad
\eqalign{
\d\bar\psi^{1,0} =&
	-\bar\ep\Dp \s
\cr
\d\psi^{0,1} =&
	-\ep\Dpp \s,
	\cr
\d\th_{z} =& -\bar\ep \nabla_{\!z}\s,
\cr
\d\th_{\bar z} =& -\ep \nabla_{\!\bar z}\s,
}
\qquad
\d \s=0.
}
For the anti-ghost multiplets we have
\eqn\ujk{
\eqalign{
\d\bar \s =& -i\bar\ep\eta -i\ep\bar\eta,
\cr
\d\eta =& \ep\Bigl(+iD -\Fr{g_s}{2}f_{z\bar z} -\Fr{1}{2}[\s,\bar \s]\Bigr),
\cr
\d\bar\eta =& \bar\ep\Bigl(-iD +\Fr{g_s}{2}f_{z\bar z} -\Fr{1}{2}[\s,\bar \s]\Bigr),
\cr
}\qquad
\eqalign{
\d\bar\chi^{1,0}_z =& -\ep H^{1,0}_z 
   + \bar\ep(\Dp v_{z}-\rd_{z}A^{1,0}),
\cr
\d\chi^{0,1}_{\bar z} =& +\bar\ep H^{0,1}_{\bar z}
   + \ep(\Dpp v_{\bar z}-\bar \rd_{\bar z}A^{0,1}).
}}
We omitted the transformation laws for the auxiliary 
fields $D$, $H^{1,0}_z$ and $H^{0,1}_{\bar z}$. 
They can easily be found from the general supersymmetry 
transformations. The worldsheet scaling dimensions for the 
fields are rearranged such that they correspond to their worldsheet 
form degree. Hence the fields $(v_{z},\l_z, f_{z\bar z})$
have dimensions $(1,1,2)$, while all the other (worldsheet scalar)
fields have zero dimension.

The two BRST supercharges are identified with the holomorphic and
anti-holomorphic differentials of $\CG$-equivariant Dolbeault
cohomology satisfying the following commutation relations \HYM
\eqn\tba{
\bs^2 = 0,\qquad \{\bs,\bbs\}=-i\CL(\s),\qquad \bbs^2=0.
}
They define the operators $\rd$ and $\bar\rd$ on the space $\CA_\CG$. 
The twisted theory is defined for arbitrary Riemann surfaces $\S$.
The $U(1)$ $\CR$-charges $(J_L, J_R)$ of the original fields
before twisting are identified with the degrees of $\CG$-equivariant 
differential forms on $\CA$. 

The localization equations are read off from the transformation rules,  
\eqn\loca{
\eqalign{
F^{0,2} &= 0,\cr
\Lambda F-\z I-\Fr{g_s^2}{2}f_{z\bar z} &= 0,
}\qquad
\eqalign{
\Dp v_{z}-\rd_z A^{1,0} &= 0,\cr
\Dpp v_{\bar z}-\bar\rd_{\bar z}A^{1,0} &= 0,
}
}
\eqn\locb{
\eqalign{
\nabla\s = 0, \qquad
d_{\!A}\s = 0,\qquad
[\s,\bar\s] = 0.
}
}
For $\S=\CP^1$ one can show that, under certain condition, 
the path integral is localized to the moduli space of holomorphic maps 
$\S\rightarrow \CM_{EH}$. The EH condition is slightly changed 
from the condition for the IR target space $\CM_{EH}$. In the 
IR however, the extra term proportional to the field strength 
on the worldsheet will become zero. As we are describing a 
topological theory, and the fixed points are not changed in 
the IR, taking the IR limit does not have any effect on the 
correlation functions of the theory. Therefore we can simply 
ignore this term.

\subsubsec{Fermion Zero-Modes}

As we have mentioned, the theory has two classically conserved ghost numbers. 
The ghost numbers are related to the $\CR$-charges of the untwisted theory 
as $(-J_L,J_R)$. Note that the BRST operators $\bs$ and $\bbs$ have 
ghost numbers $(1,0)$ and $(0,1)$ respectively. The basic bosonic fields 
$v$ and $A$ have vanishing ghost numbers. Furthermore, we find the 
following ghost numbers for the fermionic fields.
$$

\matrix{
\bar\eta     & \eta                & \p^{0,1}     & \bar\p_+^{1,0} \cr
\chi^{1,0}_z & \chi^{0,1}_{\bar z} & \th_{\bar z} & \th_z          \cr
(1,0)        & (0,1)               & (-1,0)       & (0,-1)        
}
$$
On the first line are the worldsheet scalars, on the second line the 
worldsheet one-forms, and on the last line their ghost numbers. 

As is well known at the quantum level these symmetries are broken due to the 
anomaly related to the index or the Riemann-Roch theorem. Basically, 
this is due to the fermionic zero-modes. So let us look in more detail 
to these zero-modes. We assume in the following that the gauge bundle on 
$X$ is always stable, that is, semi-stability implies stability. 
In this case, there are no covariantly constant adjoint scalar sections. 
Therefore, all the fields that are scalars on the $X$ have no 
zero-modes. These include the fermionic fields $\eta$, $\bar\eta$ and $\th$. 
Therefore the remaining fermions that may have zero-modes are the 
one-forms on $X$. As we see above, there is a pair of worldsheet scalars, 
and a pair of worldsheet one-forms. The number of covariantly constant 
adjoint valued one-forms on $X$ equals the complex dimension $n$ of the 
moduli space $\CM_{EH}$ of bundles. 
Therefore, the total ghost number anomaly for $\S$ a genus $g$ worldsheet 
is $n(1-g)$, for both the ghost numbers. This means that in order to have 
a nonvanishing correlation function $\langle\Pi\CO_a\rangle$, the total 
ghost numbers of the observables $\CO_a$ should be equal to this number.

\subsubsec{Observables and Correlation Functions}

The observables of the $A$-model are easy to construct. 
We begin with observables to be inserted on a point in $\S$.
By definition those observables
are $\CG$-equivariant differential forms on the space $\CA$ 
of all connections. Those observables
generate cohomology rings of the moduli space of EH connections
via restriction and reduction. Equivalently those observables
flow to the usual observables of the non-linear sigma model in
the infrared limit.

{}From $\d\s=0$ we see that an arbitrary
$G$-invariant polynomial $P(\s)$ of $\s$ with degree
$r$ is an observables. It corresponds to an equivariant $2r$-form,
(more precisely an $(r,r)$-form). The other observables
can be obtained by the usual descent procedure. 
Equivalently we may use the universal
bundle to construct those observables.
{}From the Bianchi identity $d_{\! A} F =0$ and the transformation
laws in \ujb, we have the following generalized Bianchi identity
\eqn\rqa{
\eqalign{
\CD \CF &=0,\cr
}
}
where
\eqn\rqb{
\eqalign{
\CD &= \bs +\bbs + \Dp + \Dpp,\cr  
\CF &= \s + i\bar\p^{1,0} + i \p^{0,1} +
F^{2,0} + F^{1,1} + F^{0,2}
}
}
We define
a generalized Chern class $\cmmib{c}_n$ by
\eqn\rqc{
\cmmib{c}_n=\Fr{(-1)^n}{(2\pi)^n n!}\tr \CF^n.
}
We expand the generalized Chern class 
as
\eqn\rqd{
\cmmib{c}_n = \sum_{p+q+r+s = 2n} \CV^{r,s}_{p,q}
}
where the upper indices denote the form degree on $X$ while
the lower indices denote the degree of the ghost number.
Now it follows from the Bianchi identity \rqa\ that we have 
the following descent equations
\eqn\descent{
(\bs +\bbs + \rd + \bar\rd) \cmmib{c}_n=0,
}
which can be written in terms of the observables as 
\eqn\rqe{
\bbs \CV^{r,s}_{p,q}
+\bs \CV^{r,s}_{p-1,q+1}
+\bar\rd \CV^{r,s-1}_{p,q+1}
+\rd \CV^{r-1,s}_{p,q+1}
=0.
}
We define
\eqn\rqf{
V_{p,q}(\a) = \int_X \a^{3-r,3-s}\wedge \CV^{r,s}_{p,q} 
}
where $\a^{3-r,3-s}\in H^{3-r,3-s}(X)$, $0\leq r,s \leq 3$ and
$0\leq p,q$. 
Then we have
equivalently
\eqn\rqg{
\bs  V^{r,s}_{p-1,q}
+\bbs V^{r,s}_{p,q-1}
=0.
}
The relation \rqe\ implies that the $Q=\bs+\bbs$ cohomology
depends on the $d$-cohomology on $X$.
From the Hodge diamond for $h^{r,s}(X)$
\eqn\hodge{
\matrix{
&&&1&&&\cr
&&0&&0&&\cr
&0&&h^{1,1}\!&&0&\cr
1&&h^{2,1}&&h^{1,2}&&1\cr
&0&&h^{1,1}\!&&0&\cr
&&0&&0&&\cr
&&&1&&&\cr
}
}
we see immediately that we can discard some of the 
$\CV^{r,s}_{p,q}$ for defining non-trivial observables.
In calculating correlation functions with these 
observables, the ghost numbers $(r,s)$ should add up 
to the total ghost number anomaly, which is $(d,d)$, 
where $d$ is the dimension of the moduli space. 
This is related to the fact that the calculation of 
the correlation function can be reduced to an 
integral over the moduli space of the corresponding 
form. As only the integral of a top form gives a 
non zero integral, we find only non zero correlation 
functions when the abovementioned condition is met.

Among other observables the equivariant K\"{a}hler form $\widetilde\Bz$ 
\bae\ plays an important role (here $\phi$ should be replaced by $\s$). 
It can be identified with the
first Chern class of a $\CG$-equivariant determinant line bundle $\CL$ over
$\CA^{1,1}$. After reduction to $\CM_{EH}$ the line bundle
becomes the determinant line bundle with the first Chern class
given by the K\"{a}hler form $\Bz$ on $\CM_{EH}$. The expectation
value $<\exp \widetilde\Bz>$ corresponds to the quantum volume
form of $\CM_{EH}$. It is also easy to introduce anti-symmetric
tensor fields on $\CM_{EH}$.  If we pick a two-dimensional homology
cycle $\g_2$ on $X$ we may construct the following observable 
\eqn\jha{
\tilde\a=\Fr{1}{4\pi^2}\int_{\g_2} \tr \Bigl(i\s F +\p^{0,1}
\wedge\bar\p^{1,0}\Bigr),
}
The $\bs$ and $\bbs$ cohomology class of $\tilde\a$ depends only
on the homology class of $\g_2$. On $\CM_{EH}$ $\tilde\a$ becomes
an element of type $(1,1)$ in the cohomology of $\CM_{EH}$.

As a last remark, note that the EH condition depends on the 
class of the K\"{a}hler form $\o$ on $X$. As we vary $\o$ the target 
space $\CM_{EH}$ may undergo a sequence of birational transformations. 
However the quantum intersection form must depend smoothly on the 
K\"ahler form, as required by the supersymmetry. The difference in 
behaviour is due to sigma-model instanton corrections, which smooth 
out these transition \GLSM.

\subsec{The $B$-Model}

We now turn to the $B$-twisting of the $\CN_{ws}=(2,2)$ model. 
In this $B$-model we set $\bar\ep_\pm=0$, while the 
$\ep_\pm$ become constant functions on the worldsheet $\S$. 
Therefore the operators $\bbs_\pm$ become the BRST charges 
for this topological model. We let the BRST generator be 
given by $\d=\ep_+\bbs_-+\ep_-\bbs_+$, satisfying $\d^2=0$. 
After twisting some fields will transform differently under
the two-dimensional Lorentz group. For example, 
the twisted $\CN_{ws}=(2,2)$ vector field contains several worldsheet 
one-forms. These are given by 
\eqn\vfib{
\eqalign{
(v_{++},\bar\eta_+,\s)\rightarrow& (v_{z},\bar\eta_z,\s_z),\cr
(v_{--},\bar\eta_-,\bar\s)\rightarrow& (v_{\bar z},\bar\eta_{\bar z},\s_{\bar z}),
}
}
Hence these fields form a multiplet of worldsheet one-forms. 
Furthermore, the other fields in this multiplet become 
worldsheet scalars. 
They are anti-ghost in the twisted model. 
For the (anti) chiral multiplet containing the target 
space gauge fields, 
we find the following anti-ghosts 
$$
\psi^{0,1}_+\rightarrow \r^{0,1}_z,\qquad
\psi^{0,1}_-\rightarrow \r^{0,1}_{\bar z}. 
$$
All the other fields become worldsheet scalars after twisting.

The BRST charges $\bbs_\pm$ satisfy the following
commutation relations
\eqn\fsdj{
\bbs_\pm^2=0,\qquad \{\bbs_+,\bbs_-\} =0.
}
as remarked earlier, they are related to the (anti-)holomorphic 
derivatives on the space $\CA$. We therefore introduce the 
following linear combinations of the BRST charges 
\eqn\bdicc{
\eqalign{
\bbs &= \bbs_+ + \bbs_-,\cr
\bs^{\!\dagger}\! & = \bbs_+ - \bbs_-.\cr
}
}
Then $\bbs$ becomes the anti-holomorphic differential of the 
$\CG$-equivariant cohomology on $\CA$, while $\bs^\dagger$ 
is the adjoint of the holomorphic equivariant differential  
$\bs$ with respect to the inner product on $\CA$.
In the infrared limit $\bbs$ and $\bs$ become the  
$\bar\rd$ and $\rd^{\dagger}$ operators on the moduli 
space $\CM_{EH}$ of EH connections. In the following 
we will work exclusively with the operator $\bbs$, as we 
are mainly interested in the $\bar\rd$-cohomology on 
the moduli space. 

It is also convenient to introduce the 
following combinations of the 'fermions' 
in the six dimensional gauge multiplet, 
\eqn\rla{
\eqalign{
\bar\p^{1,0} &=\bar\p^{1,0}_+ + \bar\p^{1,0}_-,\cr 
\c^{0,2} &= *\left((\bar\p^{1,0}_+ -\bar\p^{1,0}_-)\wedge\o^{0,3}
\right)
}
}
Note that $\chi^{0,2}$ could also be identified with a $(-1,0)$ form 
or vector, using the metric instead of $\o^{0,3}$. 

We have the following BRST transformation laws 
for the basic fields coming from the matter fields, 
\eqn\kxa{
\eqalign{
 \bbs A^{1,0} 
	=& i\bar\psi^{1,0},
\cr
 \bbs A^{0,1} 
	=& 0,
\cr
}\qquad
\eqalign{
\bbs\bar\psi^{1,0} =& 0,
\cr
\bbs\bar\chi^{0,2} =& F^{0,2},
\cr
}
}
For the one-forms from the vector field we find 
\eqn\kxb{
\eqalign{
\bbs v_{z} &=  i\bar\eta_z,\cr
\bbs v_{\bar z} &= i\bar\eta_{\bar z},\cr
}\qquad
\eqalign{
\bbs \s_z &=   -i\bar\eta_z,\cr
\bbs \s_{\bar z} &= -i\bar\eta_{\bar z},\cr
}\qquad
\eqalign{
\bbs\bar\eta_{z}
     &= 0,
\cr
\bbs\bar\eta_{\bar z} 
 	 &=0.
}
}
And finally for the anti-ghosts 
\eqn\kxc{
\eqalign{
\bbs\r^{0,1}_z &=
	-\Dpp \s_z 
	-\Dpp v_{z} +\rd_z A^{0,1},
\cr
\bbs\r^{0,1}_{\bar z} &=
	-\Dpp \s_{\bar z}
	-\Dpp v_{\bar z} +\rd_{\bar z}A^{0,1},
\cr
\bbs\eta_{+}
     =&(\L F-\z I)-\Fr{1}{2}[\s_z,\s_{\bar z}]-\Fr{1}{2}f_{z\bar z}
	-\nabla_{\!z} \s_{\bar z},
\cr
\bbs\eta_{-} 
	=&(\L F-\z I)+\Fr{1}{2}[\s_z,\s_{\bar z}]+\Fr{1}{2}f_{z\bar z}
	-\nabla_{\!\bar z}\s_z.
\cr
}
}

{}From these transformations we read off the  following
fixed point equations (here we used both the $\bbs_\pm$ 
transformation rules)
\eqn\kxd{
\eqalign{
F^{0,2}=0,\cr
\L F-\zeta I=0,\cr
\nabla_{\!\bar z}  \s_{z}=0,\cr
f_{z\bar z} + [\s_z,\s_{\bar z}]=0,\cr
}
}
and
\eqn\kxe{
\eqalign{
\Dpp \s_z= \Dpp \s_{\bar z}=0,\cr
\Dpp v_{z}  -\rd_z  A^{0,1}=0,\cr
\Dpp v_{\bar z} -\rd_{\bar z} A^{0,1}=0,\cr
}
}
Note that the last two equations in \kxd\ are Hitchin's 
self-duality equations in two dimensions \Hitchin. 
On a cylinder or $\CP^1$
these equations have no non-trivial solutions.\foot{
If we consider a Riemann surface $\S$ 
with $genus(\S) \geq 1$, the moduli space of the Hitchin
equations may play an important role.}
Thus $f_{z\bar z}=\s_z=0$. Then the connection $v_{z}$
is flat and can be gauge transformed away. What
we are left with from the above equations are
\eqn\kxf{
\rd_z A^{0,1}=
\rd_{\bar z} A^{0,1}=0.
}
Thus the path integral is localized to a copy of the moduli space
$\CM_{EH}(X)$ of EH connections on $X$.

The action functional for the $B$-model can be written in the form 
\eqn\cyab{
\eqalign{
S(e^2) = \Fr{1}{e^2}\bbs V + \Fr{1}{e^2} W,
}
}
modulo terms which vanish by the fermion equations of motion. 
The precise form for $V$ and $W$ is given in  Appendix B. 
We introduced a coupling constant $e$ for convenience later. 
The part $W$ comes from the holomorphic potential; it is invariant 
under the BRST symmetry generated by $\bbs$, although it is not exact. 
We may now follow the standard recipe for the $B$-model as put forward 
in \Wittenmirror. The correlation functions of the theory are identified
with periods of differential forms on $\CM_{EH}$.

\subsubsec{Fermion Zero-Modes}

As for the $A$-model, also the $B$-model has two classically conserved 
ghost numbers, given by the $\CR$-charges $(J_L,J_R)$. So the BRST 
operators have ghost number $(1,0)$ and $(0,1)$ respectively. We will 
consider here only the total ghost number $\Fr{1}{2}(J_L+J_R)$. 
The bosonic fields $v$ and $A$ again have vanishing ghost number. 
For the fermions, the worldsheet scalars $\eta_\pm$ and one-form 
$\r^{0,1}$ have ghost number $1$, while the worldsheet scalars 
$\bar\p^{1,0}$, $\chi^{0,2}$ and the one-form $\bar\eta$ have 
ghost number $-1$. 

Again, there is an anomaly related to the index. 
As for the $A$-model we assume that the gauge bundle on $X$ is always stable, 
so that the adjoint scalars on $X$ $\eta$ and $\bar\eta$ have no zero-modes. 
we remain with possible zero-modes for the adjoint forms. Note that the 
forms $\bar\p^{1,0}$ and $\chi^{0,2}$ have the same number of zero-modes 
on $X$, as they can be related by using $\o^{3,0}$. The number of 
covariantly constant adjoint-valued one-forms on $X$ is the complex dimension 
$n$ of the moduli space of bundles. Here it is essential that the condition 
$c_1(\CM)=0$ is met. Otherwise, the number of $\r$ zero-modes and $\bar\p$ 
and $\chi$ zero-modes would be different. This would not even lead to an 
acceptable quantum theory, as this would mean that the fermion determinant 
is not real. The total ghost number anomaly for $\S$ a genus $g$ worldsheet 
is $w = 2n(1-g)$. This means that in order to have a nonvanishing 
correlation function, the total ghost number of the observables should be 
equal to this number. 

With the assumption for the bundle on $X$ that semi-stability implies 
stability, we find that there are no zero-modes on $X$ for the adjoint 
scalars $\s$, $\eta_\pm$ and $\bar\eta$. By going to the infrared 
theory, we may therefore disregards these fields completely 
in this case. We then only remain with the one-forms on $X$. 
In the more general case when there are strictly semi-stable 
bundles, the situation becomes much harder to analyze. 
We will not deal with this situation in this paper.

\subsubsec{Some Observables}

Now we consider the observables of the $B$-model. 
We will only be concerned with situations where semi-stability 
implies stability, that is there are no strictly semi-stable bundles. 
We also restrict to the case of genus zero. 
As remarked above, we can therefore disregard all the scalars of the theory, 
while also the worldsheet dependence is trivial. 
Therefore, we will look only at the sector that is left, and replace the 
worldsheet by a point. 

The BRST transformation laws together with the Bianchi identity
$d_{\! A} F=0$ imply the following generalized Bianchi identity
\eqn\gbianki{
(\bbs+ \Dp + \Dpp)\left(i \bar\p^{1,0} + F^{2,0} + F^{1,1} + F^{0,2}\right)=0.
}
We remark that the above relation is part of the generalized
Bianchi identity \rqa\ of the $A$-model. Adopting the same procedure
as for the $A$-model we have the following partial descent equations
\eqn\rka{
\bbs \CV^{r,s}_{0, q} + \bar\rd \CV^{r,s-1}_{0,q+1}
+ \rd \CV^{r-1,s}_{0,q+1} =0.
}
Thus we can construct the following observables 
\eqn\rkb{
V_{0,q} = \int_X \a^{3-r,3-s}\wedge \CV^{r,s}_{0,q},
}
satisfying $\bbs\CV_{0,q}=0$.
Then  $V_{0,q} \in H^q_{\bar \cmmib{s}} (\CA, \bigwedge^0 T^{1,0}\CA)
\equiv H^{0,q}_{\bar \cmmib{s}} (\CA)$.
Note that the $\bbs$-cohomology is the Dolbeault cohomology on $\CA$.
An interesting observable is  
\eqn\vfa{
V_{0,3} = \Fr{i}{48\pi^3}
\int_X \o^{0,3}\wedge\tr\Bigl(\bar\p^{1,0}\wedge\bar\p^{1,0}\wedge\bar\p^{1,0}\Bigr).
}
It expresses the 
anti-holomorphic 3-form on the moduli space. 
Another intersting observable is 
\eqn\vfb{
V_{0,1} = \Fr{i}{8\pi^2}\int_X \a^{1,2}\wedge\tr\Bigl(\bar\p^{1,0}\wedge F^{1,1}\Bigr).
}
The $(1,2)$-form $\a^{1,2}$ parametrizes a deformation of the complex 
structure on the Calabi-Yau. 
It was proposed to define some special coordinates in the generalized 
mirror symmetry conjecture of \Vafa\ (or rather its complex conjugate).

To have a well-defined $B$-model we need to find observables
corresponding to  elements $V_{-p,q}$ of the Dolbeault cohomology 
$H^q_{\bar \cmmib{s}} (\CA, \bigwedge^p T^{1,0}\CA)$ with $p\neq 0$.
The natural field to use to construct observables having $p\neq0$ 
is $\chi^{2,0}$. 
However the transformation law $\bbs\chi^{0,2}=F^{0,2}$ in \kxa\
implies that there are no such observables.
We do however have $\bbs\chi^{0,2}=0$ at the fixed point locus to 
which the path integral is localized.
For example the candidate for the marginal operator $V_{-1,1}$ generating
the complex structure deformation of the moduli space $\CM_{EM}$ is
\eqn\rkc{
V_{-1,1} = \Fr{1}{8\pi^2}\int_X \a^{2,1}\wedge
  \tr\Bigl(\bar\p^{1,0}\wedge \chi^{0,2}\Bigr),
}
where $\a^{2,1}\in H^{2,1}(X)$.
We then have 
\eqn\rkd{
\bbs V_{-1,1} = -\Fr{1}{8\pi^2}\int_X \a^{2,1}\wedge
  \tr\Bigl(\bar\p^{1,0}\wedge F^{0,2}\Bigr).
}
Thus $V_{-1,1}$ certainly reduces to an element of 
$H^1_{\bar \cmmib{s}} (\CM_{EH},  T^{1,0}\CM_{EH})$.
Following \Wittenmirror\ one may try to add $V_{-1,1}$ 
to the action $S(e^2)$ and modify the $\bbs$ transformation law
in a suitable way, such that the total action deformed by this 
observable is invariant under $\bbs$. The problem with this 
approach however is that the condition $F^{0,2}=0$ is
not the equation of motion of any field.
The resolution of this will involve a deformation to holomorphic
Chern-Simons theory.

\subsubsec{Mapping to Open String Field Theory of the $B$-Model on $X$}

Our starting point is the observation that $\CW(A^{0,1})$ is invariant
under the BRST transformation of our $B$-model, since 
$\bbs A^{0,1}=0$ and $\CW$ depends only on $A^{0,1}$. 
Thus we can regard $\CW$ as an "observable" 
in our $B$-model and consider the following generalized action 
functional\foot{The holomorphic Chern-Simons form 
is not invariant under large gauge transformations, but
transforms only by integral periods of the
integral periods of the 3-form $\o^{3,0}$ \Thomas. 
We will not concern ourselves here with this subtlety.}
\eqn\rya{
\eqalign{
S^\pr(e^2) =&-\Fr{ik}{8\pi^2}\int_X\tr \o^{3,0}\wedge
\Bigl(A^{0,1} \wedge\bar\rd A^{0,1} 
+\Fr{2}{3}A^{0,1}\wedge A^{0,1}\wedge A^{0,1}\Bigr)\cr
&
+\Fr{1}{e^2}\bbs V +\Fr{1}{e^2}W.
}
}
Now the condition $F^{0,2}=0$ may occur
by the $A^{0,1}$ equation of motion.
Such Chern-Simons like observables were also considered in \Baulieu, 
but in the theory at one dimension higher.

As noted above, we want to make sense out of the action functional 
deformed by the 'observable' \rkc. Thus we consider the following more 
general action functional, including both the deformation 
above and the deformation by \rkc, 
\eqn\rya{
\eqalign{
S^\ppr(e^2, t^\a) =
&-\Fr{ik}{8\pi^2}\int_X\tr \o^{3,0}\wedge
\Bigl(A^{0,1} \wedge\bar\rd A^{0,1}
+\Fr{2}{3}A^{0,1}\wedge A^{0,1}\wedge A^{0,1}\Bigr)
\cr
&-\Fr{k}{8\pi^2}t^\a \int_X\o^{3,0}\wedge
 \m_\a\wedge\tr\left(\bar\p^{1,0}\wedge\chi^{0,2}\right)
+\Fr{1}{e^2}\bbs V+\Fr{1}{e^2}W.
}
}
Here the $\m_\a \in H^1(X, T^{1,0}X)$ ($\a=1,\cdots,h^{2,1}(X)$) 
form a basis. Note that the vector index of $\m_\a$ should be 
contracted in the action above. 
The above action functional is invariant under the following
modified transformation laws (compare with \kxa)
\eqn\rlbk{
\eqalign{
\bbs   A^{1,0} &= i \bar\p^{1,0},\cr
\bbs   A^{0,1} &= it^\a\m_\a\bar\p^{1,0},\cr
}\qquad
\eqalign{
\bbs\bar\p^{1,0}&=0,\cr
\bbs\chi^{0,2} &= F^{0,2}. 
} 
}
Note that we still have $\bbs^2=0$.\foot{We obviously have 
$\bbs^2 A^{1,0}=\bbs^2 A^{0,1}=0$, while 
$\bbs^2 \chi^{0,2} = i t^\a\m_\a \Dpp \bar\p^{1,0}$.
However the latter is closed on shell, which is good enough.
We can also make the algebra being closed off-shell
by introducing an auxiliary field $H^{0,2}$, i.e., 
$\bbs\chi^{0,2}=F^{0,2} - H^{0,2}$ and 
$\bbs H^{0,2} = i t^\a\m_\a \Dpp \bar\p^{1,0}$.
}
We note that the above perturbation is the variation of  
complex structure on $\CA$ induced by the variation of complex
structure on the Calabi-Yau $X$, i.e.,
\eqn\jka{
\bbs \rightarrow \bbs + \hat\m^i\bs_i,
}
where $\hat\m^i\in \O^{1}(\CA, T^{1,0}\CA)$.
This relates very elegantly to the fact that adding 
$t^\a\int_X\o^{3,0}\m_\a
\wedge\tr\left(\bar\p^{1,0}\wedge\chi^{2,0} \right)\in \O^{1}(\CA, T^{1,0}\CA)$
to the action functional $S^\pr(e^2)$ generates a marginal deformation
corresponding to the variation of complex structure on $\CA$!

Now we take $e^2 \rightarrow 0$ in $S^\ppr(e^2,t^\a)$
to see that the path integral (the partition function)
is localized to the moduli space of stable bundles. The fermionic
zero-modes $(\bar\p^{\bar\imath}, \chi^i)$ of $(\bar\p^{1,0},\chi^{2,0})$ 
are identified\foot{This is modulo the gauge
symmetry.} as  $\bar\p^{\bar\imath}\in H^{1,0}(\End(E),\Dp)$
and $\chi_i \in H^{0,2}(\End(E),\Dpp) \simeq H^{0,1}(\End(E),\Dpp)$.
Consequently the partition function for the action functional
$S^\pr(e^2,\t^\a)$ is identified with the generating functional
of the original $B$-model correlation functions of the marginal
vertex operators
\eqn\jka{
Z^\ppr = \int \CD(Bose) \CD(Fermi) e^{-S^\ppr(e^2,t^\a)}
= \left<\exp\left(kt^\a V^{-1,1}_\a\right)\right>_B.
} 
Now following the standard argument for the $B$-model \Wittenmirror\
we should have 
\eqn\jka{
Z^\ppr \sim \int_{\CM_{EH}}\O\wedge \rd_{i_1}\ldots\rd_{i_d}\O
}
where $\O$ is the holomorphic $d$-form on the moduli space $\CM_{EH}$.

Finally we 
regard the action functional $S^\ppr(e^2,t^\a)$ in \rya\
as a BRST-exact
deformation of a theory defined by the following
action functional 
\eqn\chcs{
I(t^\a) =-\Fr{ik}{8\pi^2}\int_X \o^{3,0}\wedge
\tr\Bigl(A \wedge\bar\rd A
+\Fr{2}{3}A\wedge A\wedge A\Bigr)
-\Fr{k}{8\pi^2}t^\a \int_X\o^{3,0}\m_\a\wedge
\tr\left(\bar\p^{1,0}\wedge  \chi^{0,2}\right).
}
Here we follow the recipe of \tdYM\HYM.
Being a BRST exact deformation we expect that the theory is independent of 
$e^2$ since we have  the same localization. Here we also assume that
there are no zero-modes of $(\bar\eta_z,\bar\eta_{\bar z}, \eta_\pm,
\r^{0,1}_z, \r^{0,1}_{\bar z})$.\foot{In such a situation
our $B$-Model reduces to a cohomological field theory on the
Calabi-Yau $3$-fold $X$.}
Then we may take an extreme limit $e^2\rightarrow \infty$
and simply drop the original $S(e^2)$ from the action $S^\ppr(e^2,t^\a)$ 
to arrive at the  equivalent action functional $I(t^\a)$.

We remark that the fermionic term in $I(t^\a)$ is crucial for 
ensuring the global fermionic symmetry \rlbk, relating
the holomorphic Cherns-Simons theory with the variation
of Hodge structure on the moduli space of stable bundles. 
The term also ensures a well-defined path integral measure 
similar to the situation in \tdYM\HYM. We view our model 
as a constructive definition of the path integral of 
holomorphic Chern-Simons theory.

So we argued that the $B$-model of our matrix string 
on a Calabi-Yau $X$ is equivalent to
Witten's open string field theory of the $B$-model \Wittencss.
Recently  Vafa suggested  such an extension of mirror
symmetry involving stable bundles on Calabi-Yau \Vafa.  
It is based on  the new understanding
of mirror symmetry as $T$-duality of $T^3$-fibered Calabi-Yau
with D-branes \SYZ. 
The extended mirror conjecture involves stable
bundles on one side and minimal Lagrangian submanifolds
on the mirror side. For Calabi-Yau $3$-folds Vafa conjectured 
mirror symmetry between Witten's open string field 
theories of the A- and $B$-models \Wittencss. 
A closely related proposal 
was suggested by Kontsevich \Kontsevich\ and Tyurin \Tyurin.
It is not clear how our approach is related to Vafa's
conjecture. We should mention that in fact Vafa gave a formula 
for the classical value of the holomorphic 3-form on the 
moduli space of bundles. This holomorphic 3-form basically 
is the observable \vfa. In our model it would not be very natural 
to calculate this observable, but rather the (quantum corrected) 
value of correlation functions involving this observable. 
This is closer to the integration of (powers of) this 3-form over 
3-cycles in the moduli space.

Our $B$-model, equivalent to the model \chcs,
computes the variation of Hodge structures on the moduli space
of stable bundles. Our $A$-model computes the quantum cohomology
ring of the moduli space of stable bundles. Following 
the well-known argument for conjectural mirror symmetry
via $\CN_{ws}=(2,2)$ superconformal theory, realized
as a sigma model with the Calabi-Yau as a target space, we may
conjecture that there are mirror pairs among our $A$- and
$B$-models involving mirror Calabi-Yau's as well as mirror
stable bundles (allow for torsion-free sheaves) along the lines 
of the mirror symmetry for higher dimensional Calabi-Yau \GMP.

\ack{
We would like to thank Robbert Dijkgraaf, Erik Verlinde and Herman 
Verlinde for useful discussions. JSP is grateful to Ralph Thomas
for sending his PhD thesis and communications.
}

\appendix{A}{Supersymmetry Transformation Rules}

In this appendix we write down the explicit $\CN_{ws}=(2,2)$ 
transformation rules. They are written in terms of the 
supersymmetry transformation 
$\d = \bar\ep_- \bs_+ +\bar\ep_+\bs_- + \ep_+\bbs_- +\ep_+\bbs_+$.

For the {\it vector multiplet} $(v_{\pm\pm},\eta_\pm,\bar\eta_\pm,\s,\bar\s,D)$ 
the supersymmetry transformations are given by 
\eqn\vector{
\eqalign{
\d v_{++} &= i\bar\ep_+\eta_+ + i\ep_+\bar\eta_+,\cr
\d v_{--} &= i\bar\ep_-\eta_- + i\ep_-\bar\eta_-,\cr
\d \s    &= -ig_s\bar\ep_+\eta_- -ig_s\ep_-\bar\eta_+,\cr
\d\bar\s &= -ig_s\bar\ep_-\eta_+ -ig_s\ep_+\bar\eta_-,\cr
\d\eta_{+}
     &=+i\ep_+ D
	-\Fr{1}{2g_s^2}\ep_+[\s,\bar\s] 
	-\Fr{1}{2}\ep_+f_\S
	-\Fr{1}{g_s}\ep_-\nabla_{\!++}\bar\s
,\cr
\d\bar\eta_{+}
     &= -i\bar\ep_+D
	+\Fr{1}{2g_s^2}\bar\ep_+[\s,\bar\s] 
  	-\Fr{1}{2}\bar\ep_+ f_\S 
	-\Fr{1}{g_s}\bar\ep_-\nabla_{\!++}\s
,\cr
\d\eta_{-} 
	&=+i\ep_-D
	+\Fr{1}{2g_s^2}\ep_-[\s,\bar\s]
	+\Fr{1}{2}\ep_-f_\S
	-\Fr{1}{g_s}\ep_+\nabla_{\!--}\s
,\cr
\d\bar\eta_{-} 
 	 &=-i\bar\ep_-D
	-\Fr{1}{2g_s^2}\bar\ep_-[\s,\bar\s] 
  	+\Fr{1}{2}\bar\ep_-f_\S
   	-\Fr{1}{g_s}\bar\ep_+ \nabla_{\!--}\bar\s
,\cr
\d D
&=
+\Fr{1}{2}\bar\ep_-\nabla_{\!++}\eta_{-}
+\Fr{1}{2g_s}\bar\ep_-[\s,\eta_+]
+\Fr{1}{2}\bar\ep_+\nabla_{\!--}\eta_{+}
+\Fr{1}{2g_s}\bar\ep_+[\bar\s,\eta_-]
\cr
&\phantom{=}
-\Fr{1}{2}\ep_-\nabla_{\!++}\bar\eta_-
-\Fr{1}{2g_s}\ep_-[\bar\s,\bar\eta_+]
-\Fr{1}{2}\ep_+\nabla_{\!--}\bar\eta_{+}
-\Fr{1}{2g_s}\ep_+[\s,\bar\eta_-].
}
}
We have defined the covariant derivatives on the worldsheet 
as $\nabla_{\!\pm\pm}=\rd_{\pm\pm}+v_{\pm\pm}$, and its 
curvature is $f_\S=[\nabla_{\!++},\nabla_{\!--}]$.

The supersymmetry transformation rules for the adjoint 
{\it chiral multiplet} $(A,\p_\pm,H)$ are given by 
\eqn\chiral{
\eqalign{
 \d A 
	=& i\bar\ep_+\p_- +i\bar\ep_-\p_+ ,
\cr
\d\p_+ =&
	+\bar\ep_+ H
	+\ep_-\nabla_{\!++}A
	+{g_s}^{-1}\ep_+[\s,A],
\cr
\d\p_- =&
	-\bar\ep_- H
	+\ep_+\nabla_{\!--}A
	+{g_s}^{-1}\ep_-[\bar\s,A],
\cr
\d H =&
	-i\ep_-\nabla_{\!++}\p_- 
	-i\ep_-[\eta_+,A]
	+\Fr{i}{g_s}\ep_-[\bar\s,\p_+] 
\cr
&	+i\ep_+\nabla_{\!--}\p_+
	+i\ep_+[\eta_-,A]
	-\Fr{i}{g_s}\ep_+[\s,\p_-].
}
}

The transformation rules for the adjoint {\it anti-chiral multiplet} 
$(\bar A,\bar\p_\pm,\bar H)$ are given by 
\eqn\achiral{
\eqalign{
 \d\bar A 
	=& i\ep_+\bar\p_- +i\ep_-\bar\p_+ ,
\cr
\d\bar\p_+ =&
	+\ep_+ \bar H
	+\bar\ep_-\nabla_{\!++}\bar A
	+{g_s}^{-1}\bar\ep_+[\bar\s,\bar A],
\cr
\d\bar\p_- =&
	-\ep_- \bar H
	+\bar\ep_+\nabla_{\!--}\bar A
	+{g_s}^{-1}\bar\ep_-[\s,\bar A],
\cr
\d \bar H =&
	-i\bar\ep_-\nabla_{\!++}\bar\p_- 
	-i\bar\ep_-[\bar\eta_+,\bar A]
	+\Fr{i}{g_s}\bar\ep_-[\s,\bar\p_+] 
\cr
&	+i\bar\ep_+\nabla_{\!--}\bar\p_+
	+i\bar\ep_+[\bar\eta_-,\bar A]
	-\Fr{i}{g_s}\bar\ep_+[\bar\s,\bar\p_-].
}
}

\appendix{B}{Action for the $B$-Model}

In this appendix we give the explicit form of the action for the $B$ 
model in terms of the action fermion $V$ and the BRST invariant term $W$, 
appearing in \cyab.

The action fermion is given by
\eqn\cyac{
\eqalign{
V =& \int_X\tr\biggl(-\chi^{0,2}\wedge * F^{2,0} 
+\Fr{i}{2}\s_z\Dp\r^{0,1}_{\bar z} +\Fr{i}{2}\s_{\bar z}\Dp\r^{0,1}_{z} 
-\Fr{i}{2}(\eta_++\eta_-)F^{1,1}
\cr
& \phantom{\int_X\tr\biggl(}
-\Fr{i}{2}[\nabla_{\!z},\Dp]\wedge\r^{0,1}_{\bar z}
-\Fr{i}{2}[\nabla_{\!\bar z},\Dp]\wedge\r^{0,1}_z 
\cr
& \phantom{\int_X\tr\biggl(}
+\Ha\Bigl(\L F-\z I+\Ha[\s_z,\s_{\bar z}]+\Ha f_{z\bar z} 
+\nabla_{\!\bar z}\s_z \Bigr)\eta_+ 
\cr
& \phantom{\int_X\tr\biggl(}
+\Ha\Bigl(\L F-\z I-\Ha[\s_z,\s_{\bar z}]-\Ha f_{z\bar z} 
+\nabla_{\!z}\s_{\bar z} \Bigr)\eta_-
\biggr).
}
}

The remaining terms in the action are given by 
\eqn\cyac{
\eqalign{
W =& \int_X\Bigl(
F^{0,2}\wedge *F^{0,2} -\Dpp\p^{0,1}_+\wedge\p^{0,1}_-\wedge\o^{3,0}\Bigr).
}
}
Note that it is invariant under the BRST symmetry $\bbs$ of the $B$-model.

\listrefs
\bye